\documentclass[journal,twoside]{IEEEtran}

\usepackage{cite}
\usepackage{amsmath,amssymb,amsfonts}
\usepackage{algorithmic}
\usepackage{graphicx}
\usepackage{textcomp}
\usepackage{epsfig} 
\usepackage{accents}
\usepackage{mathtools}
\usepackage{color}
\usepackage{balance}
\usepackage{verbatim}

 \usepackage{amsthm}

\usepackage{enumitem}

\usepackage{hyperref}
\usepackage{tikz,pgfplots,pgfplotstable,pgffor,pgfmath,tikz-3dplot}
\usepackage{tikz-dimline}

\newtheorem{theorem}{Theorem}
\newtheorem{assumption}{Assumption}

\newtheorem{remark}{Remark}
\newtheorem{lemma}{Lemma}

\begin{document}


\title{Barrier Integral Control for Global Asymptotic Tracking of Uncertain Nonlinear Systems under State and Input Constraints}

\author{Christos K. Verginis, \IEEEmembership{Member, IEEE}
\thanks{The author is with the Division of Signals and Systems, Department of Electrical Engineering, Uppsala University. 
(e-mail: christos.verginis@angstrom.uu.se). }}

\maketitle

\begin{abstract}                          
This paper addresses the problem of asymptotic {tracking} for high-order control-affine MIMO nonlinear systems with unknown dynamic terms  subject to {input} and transient state constraints. We introduce Barrier Integral Control (BRIC), a novel algorithm designed to confine the system's state within a predefined funnel, ensuring adherence to the transient state constraints, and asymptotically drive it to a given reference trajectory from any initial condition. The algorithm leverages the innovative integration of a reciprocal barrier function and error-integral terms, featuring smooth feedback control. {We further develop an extension of the algorithm, entailing continuous feedback, that uses a reference-modification technique to account for the input-saturation constraints.}
Notably, BRIC operates without relying on any information or approximation schemes for the (unknown) dynamic terms, which, unlike a large class of previous works, are not assumed to be bounded or to comply with globally Lipschitz/growth conditions. Additionally, the system's trajectory and asymptotic performance are decoupled from the uncertain model, control-gain selection, and initial conditions. 
Finally, comparative simulation studies validate the effectiveness of the proposed algorithm.

\end{abstract}

\section{Introduction} \label{sec:intro}
Controlling systems with uncertain dynamics has been a major research topic in control theory. It presents significant challenges due to uncertainties stemming from modelling inaccuracies, environmental variations, and external disturbances. 
To maximize the robustness to system uncertainties, the goal is to ultimately minimize a stabilization error and achieve a satisfying transient behaviour while reducing  the dependency on prior knowledge of system dynamics as well as {complying with input-saturation constraints}. 


Adaptive control methods are widely used to tackle
systems with uncertain nonlinear dynamics, establishing however only ``practical" stability results and/or adopting limiting assumptions.
The first category entails convergence of the control error to a residual set around zero \cite{gutierrez2023interval,ge2002adaptive}; the size of that set usually depends on the dynamics of the system \cite{gutierrez2023interval} or residual errors of dynamic-approximation schemes \cite{ge2002adaptive} and its shrinkage requires selection of excessively large gains.  
{On the other hand, robust integral control constitutes a notable class of adaptive methods that
 has focused on the asymptotic regulation of uncertain systems and motivates the results of this paper. However, such asymptotic results are typically achieved at the cost of adopting limiting assumptions on the system dynamics.  
These assumptions include globally Lipschitz conditions, knowledge of upper bounding functions/constants of the dynamic terms \cite{xudong1999asymptotic,khalil2002universal,yu2012robust,aguilar2014nonlinear,psillakis2016integrator,taleb2015adaptive,jiang2001robust,patre2006asymptotic,zhao2025beyond}, 
vanishing drift terms at the desired setpoint \cite{yu2012robust,li2018control,zhang2019adaptive}, 
SISO systems \cite{jiang2001robust,psillakis2016integrator,aguilar2014nonlinear},
 and/or unit control-input matrices  \cite{jiang2001robust}.}
Similarly, previous works that guarantee asymptotic or finite-time trajectory tracking consider known parts of the dynamic terms \cite{patil2021exponential}, linear parametrizations of the dynamics, where the uncertainty is restricted to constant or bounded time-varying terms \cite{zhang2024global,zhao2021adaptive,xian2017continuous,song2023prescribed}, 
bounded drift terms that are compensated by control-gain tuning \cite{patil2022adaptive,patre2006asymptotic}, or a priori available data \cite{verginis2023non}. 


The aforementioned assumptions are removed by a class of works that develop funnel or Prescribed Performance control algorithms \cite{berger2021funnel,bechlioulis2008robust}. These algorithms use high-gain feedback to  guarantee the evolution of the control error in a pre-defined funnel, establishing prescribed transient and steady-state properties, without using any information on the system dynamics. However, \textit{asymptotic} convergence guarantees would require the funnel to converge to zero (as, e.g., in \cite{li2018control,lee2019asymptotic}), creating numerical ill-conditioning in the implementation of the control algorithm. 
Asymptotic tracking without zero-converging funnels requires either linear systems \cite{berger2024asymptotic} or limiting the uncertainty to constant model parameters \cite{karayiannidis2012model}. Our previous works \cite{verginis2020asymptotic,verginis2023asymptotic} established asymptotic tracking for 2nd-order nonlinear systems with unknown dynamics, resorting however to \textit{discontinuous} control laws that are not applicable in practice. Similarly, \cite{zhou2023asymptotic} develops a control algorithm that converges to an undefined expression and requires neural-network approximations that are only \textit{locally} valid. 

{Additionally, a critical property associated with control of real-world systems is that of input-saturation, especially with high-gain funnel-control methods that cause transient spikes in the control input. Notable works have been developed towards this direction, by using switching controllers, modification of the reference trajectory or the funnel, or assuming large enough saturation bounds \cite{trakas2023robust,bikas2024prescribed,fotiadis2023input}. 
Nevertheless, these works establish funnel containment without asymptotic guarantees and focus mostly on SISO systems.}

This paper considers the {trajectory-tracking} problem for high-order MIMO nonlinear systems with unknown dynamic terms {under input and state constraints}.
We propose Barrier Integral Control (BRIC),  a novel algorithm that guarantees the evolution of the state in a pre-defined funnel and its asymptotic convergence to a given reference trajectory from all initial conditions. {The reference trajectory is considered to belong to a certain class of trajectories that satisfy a square-integrability criterion along the system dynamics.}
BRIC comprises a reciprocal barrier function and two error-integrator terms, which guarantee the boundedness of the error in the funnel and its asymptotic convergence to zero, respectively.  
{We further develop an extension of the introduced BRIC algorithm that employs a reference-modification scheme to account for input-saturation constraints. Unlike previous related works, the input-constrained BRIC takes into account the MIMO structure of the system with non-diagonal control-input matrices.}
The performance of the closed-loop system is independent from the system's dynamics, the initial conditions, and the selection of the control gains. Unlike a large majority of related works, we do not impose growth conditions, linear parametrizations, or global boundedness on these  terms, and we do not employ any schemes, such as neural networks, to approximate them. 
{Moreover, unlike previous results that achieve asymptotic tracking using discontinuous feedback \cite{verginis2020asymptotic,verginis2023asymptotic,taleb2015adaptive}, the nominal and input-constrained BRIC versions use smooth and continuous feedback, respectively. The algorithms of \cite{verginis2020asymptotic,verginis2023asymptotic,taleb2015adaptive} cannot be trivially modified into continuous/smooth versions to solve the asymptotic tracking problem, even for the restricted class of reference trajectories considered in this work, especially when taking into account input-saturation constraints.}
This paper extends our preliminary work \cite{verginis2024asymptotic}
by considering a larger class of high-order systems, { taking into account input constraints}, and removing a number of assumptions on the dynamics. 

The rest of the paper is organized as follows. Sec. \ref{sec:PF} provides the problem formulation. Sec. \ref{sec:main results} and \ref{sec:main results input sat} illustrate the nominal and input-constrained BRIC, respectively. Finally, Sec. \ref{sec:sims} provides simulation results and Sec. \ref{sec:conclusion} concludes the paper.

\section{Problem Formulation} \label{sec:PF}

We consider the problem of asymptotic tracking of MIMO systems of the form 	
\begin{subequations} \label{eq:dynamics}
\begin{align}
\dot{x}_i &= x_{i+1}, \ \ i\in\{1,\dots,k-1\} \label{eq:dynamics xi} \\
\dot{x}_k &= F(x,z,t) + G(x,z,t)u \label{eq:dynamics xk} \\
\dot{z} &= F_z(x,z,t) \label{eq:dynamics z} 
\end{align}
\end{subequations}
where $z \in\mathbb{R}^{n_z}$, $x\coloneqq [x_1^\top,\dots,x_k^\top]^\top \in \mathbb{R}^{kn}$, with $k\geq 2$, $x_i \coloneqq [x_{i_1},\dots,x_{i_n}]^\top \in \mathbb{R}^n$, for all $ i\in\{1,\dots,k\}$, {are} the system's states, $u\in\mathbb{R}^n$ is the control input, and $F:\mathbb{R}^{kn+n_z}\times \mathbb{R}_{\geq 0} \to \mathbb{R}^n$, $F_z : \mathbb{R}^{kn+n_z}\times \mathbb{R}_{\geq 0} \to \mathbb{R}^{n_z} $  $G:\mathbb{R}^{kn+n_z} \times \mathbb{R}_{\geq 0} \to \mathbb{R}^{n\times n}$ are \textit{unknown} vector fields. 
The terms $F(\cdot)$ and $G(\cdot)$ entail modelling uncertainties, unknown parameters and nonlinearities, as well as external time-varying disturbances, and can model a large variety of systems \cite{Khalil}.
We further assume that $x$ is available for measurement, whereas $z$ is not. In fact, $z$ evolves subject to the system's \textit{internal dynamics} $F_z(\cdot)$, which represent unmodeled dynamic phenomena that  affect the closed{-}loop response. 

The considered problem is the design of a control protocol that outputs a smooth feedback controller $u:\mathbb{R}^{kn}\times \mathbb{R}_{\geq 0}\to\mathbb{R}^n$ to accomplish two control objectives. The primary control objective is global asymptotic {tracking of a trajectory $x_\textup{d}:\mathbb{R}_{\geq 0} \to \mathbb{R}^n$, i.e., 
$\lim_{t\to \infty} e(t) = 0$, 
where $e \coloneqq  [e_1^\top, e_2^\top, \dots, e_k^\top]^\top \coloneqq  x - \bar{x}_\textup{d}$, 
$\bar{x}_\textup{d} \coloneqq [x_\textup{d}^\top, \dots, (x_\textup{d}^{(k-1)})^\top ]^\top$,
and $e_i \coloneqq [e_{i_1},\dots,e_{i_n}]^\top \coloneqq x_i - x_{\textup{d}}^{(i-1)} \in \mathbb{R}^n$ for all $i\in\{1,\dots,k\}$; $\bar{x}_\textup{d}(t)$ is assumed to be bounded for all $t\geq 0$.}
The secondary control objective is the establishment of a predefined convergence rate for the transient phase of $e(t)$, i.e., 
\begin{align} \label{eq:transient conv rate}
\|e(t)\| \leq A \exp(- L t) + B
\end{align}
for \textit{pre-defined} positive constants $A$, $L$, and $B$.
Such an establishment is a standard result of various funnel-based  works \cite{berger2021funnel,bechlioulis2008robust}, where $B$ depends on the final value of the funnel functions. Therefore,  convergence of $e(t)$ to zero requires tuning of $B$ to arbitrarily small values, which can create issues in practical implementations (see Section \ref{sec:intro}). 
In this work, we guarantee that $\lim_{t\to \infty} e(t) = 0$ independent of $B$ and \eqref{eq:transient conv rate} dictates only the transient behaviour of $e(t)$.
In order to solve the aforementioned problem, we consider the following assumptions on the system dynamics: 
{\begin{assumption} \label{ass:f(x,t) cont + bounded}
	The maps $(x,z) \mapsto F(x,z,t) : \mathbb{R}^{kn+n_z}\to\mathbb{R}^n$, $(x,z) \mapsto G(x,z,t) : \mathbb{R}^{kn+n_z}\to\mathbb{R}^n$, $(x,z) \mapsto F_z(x,z,t) : \mathbb{R}^{kn+n_z}\to\mathbb{R}^{n_z}$ are {locally} Lipschitz continuous for each fixed $t \geq 0$, uniformly in $t$, and the maps $t\mapsto F(x,z,t):\mathbb{R}_{\geq 0} \to\mathbb{R}^{n}$ and $t\mapsto G(x,z,t):\mathbb{R}_{\geq 0} \to\mathbb{R}^{n}$ are continuous and uniformly bounded for each fixed $(x,z)\in \mathbb{R}^{kn+n_z}$.   
\end{assumption}
\begin{assumption} \label{ass:g(x,t) pd}
The matrix $ \widetilde{G}(x,z,t) \coloneqq G(x,z,t) + G(x,z,t)^\top $
is {positive definite, i.e.,
${\lambda}_{\min}( \widetilde{G}(x,z,t)  )  > 0$, $\forall (x,z,t) \in \mathbb{R}^{kn+n_z} \times \mathbb{R}_{\geq 0}$, where ${\lambda}_{\min}( \widetilde{G}(x,z,t))s$ denotes its unknown minimum eigenvalue.}
\end{assumption}
\begin{assumption} \label{ass:internal dynamics}	
	There exists a sufficiently smooth function $U_z:\mathbb{R}^{n_z}\to\mathbb{R}_{\geq 0}$ and class $\mathcal{K}_\infty$ functions $\underline{\gamma}_z(\cdot)$, $\bar{\gamma}_z(\cdot)$, $\gamma_z(\cdot)$ such that
	$\underline{\gamma}_z(\|z\|) \leq {U_z}(z) \leq \bar{\gamma}_z(\|z\|)$, and
	\begin{equation*}
		{\left(\frac{\partial U_z}{\partial z}\right)^\top F_z(x,z,t)} \leq -\gamma_z(\|z\|) + \pi_z(x,z,t), 
	\end{equation*}
	where $(x,z)\mapsto \pi_z(x,z,t) :\mathbb{R}^{kn+n_z} \to \mathbb{R}_{\geq 0}$ is continuous and class $\mathcal{K}_\infty$ for each  $t \geq 0$, and $t \mapsto \pi_z(x,z,t): \mathbb{R}_{\geq 0}  \to \mathbb{R}_{\geq 0}$ is uniformly bounded for each  $(x,z)\in \mathbb{R}^{kn+n_z}$.
\end{assumption}
\begin{assumption} \label{ass:F = 0}
{
There exist unknown constants $F_\textup{d} \in \mathbb{R}^n$, $\bar{W}\geq 0$, and an unknown matrix $G_\textup{d} \in \mathbb{R}^{n \times n}$ with positive definite symmetric part $G_\textup{d}+G_\textup{d}^\top$  
such that, for all $\mathsf{z}:\mathbb{R}_{\geq 0}\to\mathbb{R}^{n_z}$ satisfying $\|\mathsf{z}(t)\| \leq \bar{\mathsf{z}}$, $t\geq 0$, with 
 a positive constant $\bar{\mathsf{z}}$, 
it holds that 
\begin{align*}
& \int_0^{\infty} \bigg( \left\|G(\bar{x}_\textup{d}(\tau),\mathsf{z}(\tau),\tau) - G_\textup{d}  \right\|^2 + \\
& \hspace{10mm} \left\|F(\bar{x}_\textup{d}(\tau),\mathsf{z}(\tau),\tau) - F_\textup{d} - x_\textup{d}(\tau)^{(k)} \right\|^2 \bigg) \textup{d}\tau \leq \bar{W}
\end{align*}}
\end{assumption}
Assumption \ref{ass:f(x,t) cont + bounded} provides mild regularity conditions for the existence of solutions of \eqref{eq:dynamics}.
{Assumption \ref{ass:g(x,t) pd} is a sufficient controllability condition, also adopted in a large variety of related works (e.g., \cite{bechlioulis2008robust,zhou2023asymptotic,berger2021funnel}). Cases of sign-definite $\widetilde{G}(\cdot)$, with known sign of ${\lambda}_{\min}( \widetilde{G}(x,z,t))$ can be trivially accommodated by using that sign in the control algorithm (such as, e.g., \cite{khalil2002universal}). Unknown signs can be handled by using the concept of Nussbaum functions \cite{yu2012robust,li2018control,chen2019nussbaum}, which, however, cannot be trivially incorporated in the proposed integrator-based algorithm and consists part of our future work.
  Next, Assumption \ref{ass:internal dynamics} is a mild condition that suggests that $z$ is input-to-state practically stable with respect to $x$, implying stable internal dynamics, and adopted in a wide variety of works \cite{khalil2002universal,yu2012robust,jiang2001robust}}.
{Finally, Assumption \ref{ass:F = 0} essentially implies that the integrand converges to zero fast enough. In practice, the simplest (but not the only) way to satisfy such a condition is $\lim_{t\to\infty}x_\textup{d}(t)^{(k)} = 0$  and $\lim_{t\to\infty} F(\bar{x}_{\textup{d}}(t),z(t),t) = F_\textup{d}$,  $\lim_{t\to\infty} G(\bar{x}_{\textup{d}}(t),z(t),t) = G_\textup{d}$ for any bounded $z(t)$. For $k=2$, for instance, this corresponds to trajectories converging to a constant setpoint and dynamics converging to constant terms along $x_\textup{d}(t)$. Intuitively, $F(\cdot)$ and $G(\cdot)$ converge to constants as $x_\textup{d}(t)$ converges to a setpoint. Note that several works that establish asymptotic stabilization properties consider stricter assumptions, such as $F(x_\textup{d},t) = 0$ for constant $x_\textup{d}$ \cite{yu2012robust,li2018control,zhang2019adaptive}. Finally, note that Assumption \ref{ass:F = 0} concerns only the reference trajectory $x_\textup{d}(t)$; for any other $x$, the terms $F(x,z,t)$, $G(x,z,t)$ are affected by the internal dynamics of $z$ as well as potential time-varying disturbances.  }

\section{Barrier Integral Control} \label{sec:main results}

To solve the global asymptotic tracking problem presented in Section \ref{sec:PF}, we propose a novel control algorithm, referred to as  Barrier Integral Control (BRIC).
More specifically, the BRIC algorithm consists of the integration of reciprocal barrier functions and adaptive control. Loosely speaking, the deployed barrier function establishes the boundedness of $x(t)$ in a compact domain, while appropriate adaptation integrals aim to compensate for the unknown upper bounds of the dynamics in this domain. In this work, we employ additional barrier terms to establish global results from any initial conditions.
We begin by defining  
 the linear filter-type errors 
\begin{align}  \label{eq:s_i}
s_i &\coloneqq [s_{i_1},\dots,s_{i_n}]^\top \notag \\
 &\coloneqq \left( \frac{\textup{d}}{\textup{d}t} + \lambda \right)^{i-1}e_1 \ = 
 \  \sum_{\ell=0}^{i-1} \left( \begin{matrix}
i- 1\\
\ell
\end{matrix} \right)
\lambda^\ell e_{i-\ell} 
\end{align}
for all $i\in\{1,\dots,k\}$ and
a positive constant $\lambda$. 
Note that \eqref{eq:s_i} can be considered as a set of stable linear filters with $s_i$ and $e_1$ being their inputs and output, respectively.  Therefore, $\lim_{t\to\infty}s_i(t) = 0$, which we aim at establishing, implies $\lim_{t\to\infty}e_\ell(t) = 0$,  for all $\ell\in\{1,\dots,i\}$,
$i\in\{1,\dots,k\}$.
{Further, by using the Laplace transform, one can easily conclude from \eqref{eq:s_i} that }
\begin{align} \label{eq:s_i+1}
s_{i+1} = \dot{s}_i + \lambda s_i, 
\end{align}	
for all $i\in\{1,\dots,k-1\}$, which will be used in the sequel. 

In order to satisfy the transient constraint \eqref{eq:transient conv rate}, we aim to bound the combined error $s_k$ in a prescribed funnel. Such a funnel further needs to attain infinite values at $t=0$ in order to accommodate the \textit{global} tracking requirement. 
Consider, therefore, functions $\phi_j:{(0,\infty)} \to (\underline{\phi}_j,\infty)$, where $\underline{\phi}_j$ are positive constants, which satisfy $\lim_{t\to 0^{+}}\phi_j(t) = \infty$ and are continuously differentiable for all $t > 0$ and $j\in\{1,\dots,n\}$.
We further require the boundedness of {$\lim_{t\to 0^+}\frac{\dot{\phi}_j(t)}{\phi_j(t)^3}$} and  $\frac{\dot{\phi}_j(t)}{\phi_j(t)^3}$ for all $t > 0$ and $j\in\{1,\dots,n\}$.
The funnel-confinement goal is to guarantee $ -\phi_j(t) < s_{k_j}(t) < \phi_j(t)$, for all {$t > 0$} and $j\in\{1,\dots,n\}$; {note that the inequality is satisfied for  $t\to 0^{+}$ by construction}.
For instance, $\phi_j(t)$ can be chosen as the exponentially decreasing functions $\phi_j(t) = \frac{1}{t}\exp(-c_{r,j} t) + \underline{\phi}_j$, where $c_{r,j} > 0$ is the convergence rate and $\underline{\phi}_j > 0$ is the final value. 
This, in turn, guarantees the establishment of the transient constraint \eqref{eq:transient conv rate} for some constants $A$, $L$, and $B$, {as shown later in Lemma \ref{lemma:s bounds}.}
We proceed now with the derivation of the BRIC algorithm. We first define the mappings 
$\Xi : \mathbb{R} \to (-1,1)$, $T:(-1,1)\to\mathbb{R}$, and $\Psi : \mathbb{R}_{> 0} \to (1,\infty)$, with
\begin{align*}
\Xi(\ast) \coloneqq & \frac{\ast}{\sqrt{\ast^2 + \kappa}}, \
 \Psi(\ast) \coloneqq & \sqrt{\frac{1}{\ast^2} + 1}, \ 
 T(\ast) \coloneqq & \frac{\ast}{1-\ast^2}
\end{align*}
for a positive constant $\kappa$, which will be used in the sequel. Note that $\Xi(\ast)$ and $T(\ast)$ are monotonically increasing whereas $\Psi(\ast)$ is monotonically decreasing. Additionally, 
$T(\ast)$ satisfies $\lim_{\ast \to \pm 1} T(\ast) = \pm \infty$ and $T(0) = 0$\footnote{Other transformation satisfying these properties could also be used, such as $T(\ast) = \ln\left( \frac{1 + \ast}{1 - \ast} \right)$.}. For future reference, we further note the inverse of $\Xi(\ast)$ and $\Psi(\ast)$ and the derivatives of $\Xi(\ast)$ and $T(\ast)$ as  
\begin{subequations} \label{eq:inverses and derivatives}
\begin{align} 
& \Xi^{-1}(\ast) = \frac{\ast\sqrt{\kappa}}{\sqrt{1-\ast^2}}, \ \  \Psi^{-1}(\ast) = \frac{1}{\sqrt{\ast^2-1}} \\
& \Xi'(\ast) = \frac{\kappa}{(\ast^2 + \kappa)\sqrt{\ast^2 + \kappa}}, \ \  
T'(\ast) = \frac{1+\ast^2}{(1-\ast^2)^2}
\end{align}
\end{subequations}

\noindent \textbf{Step I}: In order for the proposed control algorithm to be independent of the initial condition, we define the transformation 
\begin{align} \label{eq:eta}
\eta_j \coloneqq \Xi(s_{k_j}) = \frac{s_{k_j}}{\sqrt{s_{k_j}^2 + \kappa}}, \ j\in\{1,\dots,n\}.
\end{align}

\noindent \textbf{Step II}: We define the continuously differentiable funnel-function transformations {$\beta_j:[0,\infty) \to [1,\bar{\beta}_j]$ as 
\begin{subequations}\label{eq:beta}
\begin{align} 
& \beta_j(t) \coloneqq \begin{cases}
1, & \textup{ if } t = 0 \\
\Psi(\phi_j) = \sqrt{ \frac{1}{\phi_j(t)^2} + 1}, & \textup{ if } t > 0 
\end{cases} \\ 
& \dot{\beta}_j(0) = \lim_{t\to 0^+} \frac{-\dot{\phi}_j(t)}{\phi_j^3(t)} 
\end{align}
\end{subequations}}
for constants $\bar{\beta}_j > 1$, $j\in\{1,\dots,n\}$. Note that the properties of $\phi_j(t)$ imply that $\beta_j(t)$ satisfy $\beta_j(t) > \beta_j(0) = 1$ for all $t > 0$, are continuously differentiable, bounded, and have bounded time derivatives for all $t\geq 0$, and $j\in\{1,\dots,n\}$. 
We further define the normalization states $\zeta_j \coloneqq \beta_j \eta_j$ as well as the
 transformation error\footnote{We note that $\Xi(\cdot)$ and $T(\cdot)$  were also used in \cite{zhao2021adaptive}, dealing with SISO systems and parametric uncertainties.}
\begin{align} \label{eq:T transf}
\chi_j \coloneqq T(\zeta_j) = \frac{\zeta_j}{1 - \zeta_j^2}, \ j\in\{1,\dots,n\}.
\end{align} 
 

\noindent \textbf{Step III}: Define the adaptation variables $\hat{d}_1 \in \mathbb{R}$, $\hat{d}_2 \in  \mathbb{R}^n$, 
and design the BRIC protocol as 
\begin{subequations} \label{eq:control algorithm}
\begin{align} 
u &= -(\mu_g  + \hat{d}_1 + \|\hat{d}_2\|^2) \beta(t) R_\Xi(s_k)  R_T(\zeta) \chi - \hat{d}_2 \label{eq:control law}\\
\dot{\hat{d}}_1 &= \mu_{d_1} \left\| R_T(\zeta) \chi \right\|^2, \ \hat{d}_1(0) > 0 \label{eq:adaptation law} \\
\dot{\hat{d}}_2 &= \mu_{d_2} \beta(t) R_\Xi(s_k) R_T(\zeta) \chi, \ \hat{d}_2(0) \in \mathbb{R}^n \label{eq:adaptation law 2} 
\end{align}
\end{subequations}
where $\chi \coloneqq [\chi_1,\dots,\chi_n]^\top$, $\zeta \coloneqq [\zeta_1,\dots,\zeta_n]^\top \in \mathbb{R}^n$,
$\beta \coloneqq \textup{diag}\{\beta_1,\dots,\beta_n\} \in \mathbb{R}^{n\times n}$, and $\mu_{d_1}$, $\mu_{d_2}$, $\mu_g$ are positive gains. Further, $R_\Xi \coloneqq \textup{diag}\{[\Xi'(s_{k_j})]_{j\in\{1,\dots,n\}}\}$, 
$R_T \coloneqq \textup{diag}\{[ T'(\zeta_j) ]_{j\in\{1,\dots,n\}}\}$ are the derivatives of $\Xi$ and $T$ respectively, given in \eqref{eq:inverses and derivatives}.

\textbf{\textit{Control design philosophy}}: We now elaborate on the philosophy of the BRIC algorithm. 
In order to achieve the asymptotic stability and exponential convergence rate objectives, the primary aim of the control algorithm \eqref{eq:eta}-\eqref{eq:control algorithm} is the containment of the signals $\zeta_j(t)$ in $(-1,1)$, for all $j\in\{1,\dots,n\}$.
{The transformation $\Xi(\ast)$ is used to establish the independence of the theoretical guarantees from the initial condition $s_k(0)$, since it holds that $\beta_j(0) = 1$ and $|\eta_j(0)| < 1$ regardless of $s_{k_j}(0)$, leading to $|\zeta_j(0)| < 1$, $j\in\{1,\dots,n\}$ for any initial condition. Although $\phi_j(t)$ are not defined at $t=0$, the control algorithm uses the well-defined $\beta_j(t)$ that satisfy $\beta_j(0)=1$. Hence, the initial control effort is proportional to $\chi_j(0) = \frac{\eta_j(0)}{1-\eta_j(0)^2}$, $j\in\{1,\dots,n\}$, i.e., it depends on the proximity of $\eta_j(0) = \frac{s_{k_j}(0)}{\sqrt{s_{k_j}(0)^2 + \kappa}}$ to one,  which can be adjusted via the normalized coefficient $\kappa$ in \eqref{eq:eta}. This is in contrast to standard Prescribed Performance Control works \cite{bechlioulis2008robust}, where the initial control effort depends on the proximity of the tracking errors to the funnel at $t=0$.}
Next, the reciprocal barrier transformation $T(\ast)$ is used in the control algorithm \eqref{eq:control algorithm} in order to guarantee that $|\zeta_j(t)| < 1$ for all $j\in\{1,\dots,n\}$ and $t > 0$. 
In particular, in view of the property $\lim_{\ast \to \pm 1} T(\ast) = \pm \infty$ of $T(\ast)$, the control algorithm aims at retaining the boundedness of $\chi(t)$, which implies the existence of a constant $\bar{\zeta} \in (0,1)$ such that $|\zeta_j(t)| < \bar{\zeta}$ for all $j\in\{1,\dots,n\}$. 
Consequently, it holds that $ -\frac{\bar{\zeta}}{\beta_j(t)} \leq \eta_j(t)\leq \frac{\bar{\zeta}}{\beta_j(t)}$,
and by using the inverse of $\Xi(\ast)$ and of $\Psi(\ast)$ and \eqref{eq:beta},
\begin{align} \label{eq:s ineq}
|s_{k_j}(t)| \leq \frac{\bar{\zeta}\sqrt{\kappa}}{\sqrt{\beta_j(t)^2-\bar{\zeta}^2}} \leq 
\frac{\bar{\zeta}\sqrt{\kappa}}{\sqrt{1-\bar{\zeta}^2}} =: \bar{s}
\end{align}
for all $j\in\{1,\dots,n\}$. 
Moreover, since $\bar{\zeta} < 1$, the left inequality of \eqref{eq:s ineq} implies $|s_{k_j}(t)| \leq \frac{\sqrt{\kappa}}{\sqrt{\beta_j(t)^2-1}}$, which, by using $\beta_j(t) = \Psi(\phi_j(t))$ from \eqref{eq:beta} and the inverse of $\Psi(\ast)$ given in \eqref{eq:inverses and derivatives}, leads to
$-\phi_j(t)  <  
s_{k_j}(t) < \phi_j(t)$ 
for  $j\in\{1,\dots,n\}$ and $t > 0$, establishing the funnel specification. 
Additionally, it holds that $|s_{k_j}(0)| < \infty$ and $|s_{k_j}(t)| < \frac{\sqrt{\kappa}}{\sqrt{\beta_j(t)^2-1}}$, which is finite for $t$ in any compact subset of $[0,\infty)$.
Further, $s_k$ in \eqref{eq:s_i} can be considered as a set of stable linear filters with inputs $s_{k_j}$ and outputs $e_{1_j}$; hence, {\eqref{eq:s ineq} implies the transient convergence \eqref{eq:transient conv rate}, as the next Lemma shows and whose proof can be found in the appendix.
\begin{lemma} \label{lemma:s bounds}
Assume that $|s_{k_j}(t)|\leq \bar{s}$, for a constant $\bar{s} > 0$ and all $t \geq 0$, $j\in\{1,\dots,n\}$. Then, there exist positive constants $\bar{e}_{i,1}$, $\bar{e}_{i,2}$, and $\lambda_0 \in (0,\lambda)$ such that, for all $i\in\{1,\dots,k-1\}$, $j\in\{1,\dots,n\}$, $t\geq 0$,
\begin{align}
|e_{i_j}(t)| \leq \bar{e}_{i,1} \exp(-\lambda_0 t) + \bar{e}_{i,2}
\end{align}
\end{lemma} 
}
Besides the transient specification, the BRIC algorithm further aims to guarantee that $\lim_{t\to\infty}e(t) = 0$, which is accomplished via the integrator terms $\hat{d}_1$ and $\hat{d}_2$. 
We are now ready to state the main result of this paper.
\begin{theorem} \label{thm:main theorem}
Let the dynamics \eqref{eq:dynamics} under Assumptions \ref{ass:f(x,t) cont + bounded}-\ref{ass:F = 0}. 
Then, for any initial condition $x(0)$, the BRIC algorithm \eqref{eq:control algorithm} guarantees the existence of  constants $\bar{\zeta}\in(0,1)$, $\lambda_0 \in (0,\lambda)$, $\bar{e}_{i,1}$, $\bar{e}_{i,2}$ such that { (i) $|s_{k_j}(t)| < \phi_j(t)$, (ii) $|e_i(t)| < \bar{e}_{i,1}\exp(-\lambda_0 t) + \bar{e}_{i,2} \frac{\bar{\zeta}\sqrt{\kappa}}{\lambda^{k-i}\sqrt{1-\bar{\zeta}^2}} $, for all $t > 0$, $j\in\{1,\dots,n\}$, $i\in\{1,\dots,k-1\}$,  (iii) $\lim_{t\to \infty} e(t) = 0$, and (iv) the boundedness of all closed-loop signals.}
\end{theorem}
\begin{remark}[\textbf{Approx.-free asymptotic tracking}]
Note that no information regarding the unknown  terms $F(\cdot)$ and $G(\cdot)$ is incorporated in the BRIC algorithm \eqref{eq:control algorithm}. These terms are further not assumed to satisfy any global boundedness/Lipschitz, or growth conditions, and no schemes are used to approximate them. 
Still, the proposed BRIC algorithm drives $e(t)$ to zero.
As illustrated in the proof of Theorem \ref{thm:main theorem}, the drift term $F(\cdot)$ is proven bounded due to the containment in the funnel \eqref{eq:s ineq} via the barrier signals $\chi_j$. This bound, unknown to the control design, is eventually compensated by the integral term $d_1(t)$ of \eqref{eq:adaptation law}, which is proven to converge to a finite value.
Similarly, the integral term $\hat{d}_2(t)$  is used to compensate for the (unknown) term $G_d^{-1} F_d$ from Assumption \ref{ass:F = 0}. 
{Integral terms similar to $\hat{d}_1(t)$ were also used in \cite{zhang2019adaptive,zhou2023asymptotic} as well as  our previous works \cite{verginis2024asymptotic,verginis2020asymptotic} to establish  asymptotic guarantees, exhibiting however, certain limitations. In particular, \cite{zhang2019adaptive} considers the simpler problem of stabilization to zero, with vanishing drift terms, i.e., $F(0,z,t) = 0$, which implies a growth condition on $F(\cdot,z,t)$; \cite{zhou2023asymptotic} develops a control law that converges to an undefined expression as $t\to\infty$ and has a local region of attraction due to the approximation of the dynamics by neural networks. Our previous work \cite{verginis2024asymptotic} was limited to a more restricted class of systems, while \cite{verginis2020asymptotic} developed a discontinuous control law for asymptotic tracking.
We stress that a continuous version of \cite{verginis2020asymptotic}'s algorithm cannot be trivially shown to solve the asymptotic tracking problem, even under Assumption \ref{ass:F = 0}, simply by adding the integrator $\hat{d}_2(t)$. The required analysis is significantly different and further complicated by the \textit{global} tracking property as well as the control-saturation constraints (handled in Sec. \ref{sec:main results input sat})}.  
\end{remark}


{
\begin{remark}[\textbf{Measurement noise}]
BRIC can also take into account measurement noise as in $\widetilde{s}_k = s_k + n(t)$, where $n(t)$ is a differentiable (almost everywhere) and bounded noise function, with $\dot{n}(t)$ satisfying a  condition similar to Assumption \ref{ass:F = 0}. Funnel specifications can then be established for $\widetilde{s}_k(t)$; if an upper bound of $n(t)$ is known, such a funnel can be adjusted to include $s_k(t)$.
Establishment of funnel specifications for the original error $e_1(t)$ could also be guaranteed by assuming, however, noise measurements only in the higher order states $x_i$, $i\in\{2,\dots,k\}$, and following a back-stepping-like design funnel methodology (see, e.g., \cite{verginis2019robust}).
\end{remark}}

\begin{proof} 
The proof follows three main steps: first, the existence of a local maximal solution $\zeta(t)$ satisfying $\zeta_j(t) \in (-1,1)$ for $j\in\{1,\dots,n\}$ and some $[0,t_{\max})$, with $t_{\max} > 0$; second, the extension of $t_{\max}$ to $\infty$; and third, the asymptotic stabilization of $e(t)$ to $0$. 

\noindent \textbf{1) Existence of a maximal solution.}\\
{We define the vector $\bar{x} \coloneqq [\zeta^\top, x^\top, z^\top, \hat{d}_1, \hat{d}_2^\top]^\top$ and the set  
$\bar{\Omega} \coloneqq (-1,1)^n \times \mathbb{R}^{(k+1)n +n_z+1}$}. 
{At $t=0$, it holds that $\zeta_j(0) = \beta_j(0) \eta_j(0) =  \eta_j(0)$, which, by construction satisfies $\eta_j(0) \in (-1,1)$, $j\in\{1,\dots,n\}$, and hence $\bar{x}(0) \in \bar{\Omega}$.}
 By combining \eqref{eq:dynamics} with \eqref{eq:control algorithm}, we obtain the closed-loop dynamics $\dot{\bar{x}} = f_{\textup{cl}}(\bar{x},t)$. 
In view of Assumption \ref{ass:f(x,t) cont + bounded}, it can be concluded that $f_{\textup{cl}}(\bar{x},t)$ is locally Lipschitz continuous in $\bar{x}$, uniformly in $t$, and continuous and uniformly bounded in $t$. Therefore, \cite[Theorem 54]{sontag2013mathematical} guarantees the existence of a local solution $\bar{x}(t)$, defined in $J \coloneqq [0,t_{\max})$, with $t_{\max} > 0$, satisfying $\bar{x}(t) \in \bar{\Omega}$, and hence $\zeta_j(t) \in (-1,1)$, for all $t\in J$ and $j\in\{1,\dots,n\}$. 

\noindent \textbf{2) Extension  of $t_{\max}$ to $\infty$.} \\
We next prove that $t_{\max} = \infty$. 
Aiming to reach a contradiction, assume that this is not the case, i.e., that $t_{\max} < \infty$. 
According to \cite[Theorem 2.1.4]{bressan2007introduction}, $t_{\max} < \infty$ implies that $\lim_{t\to t_{\max}}\|\bar{x}\| = \infty$; this can be proven to be equivalent to the existence of a nonempty set $\mathcal{Z}_{\infty} \subseteq \{1,\dots,n\}$ and positive constants $\bar{\zeta}_j < 1$, $j\in \{1,\dots,n\} \backslash \mathcal{Z}_\infty$ 
such that 
$\lim_{t\to t_{\max}} |\zeta_j(t)|  = 1$ for all $j\in \mathcal{Z}_\infty$ and $|\zeta_j(t)| \leq \bar{\zeta}_j < 1$ for all  $j\in \{1,\dots,n\} \backslash \mathcal{Z}_\infty$ and $t\in J$.
For  $\zeta_j$, $j\in \mathcal{Z}_\infty$, the continuity of the solution $\bar{x}(t)$ implies that, for any constant $\mathsf{y} \in (|\zeta_j(0)|, 1)$, there exist time instants $\tau_j \in (0,t_{\max})$ such that $\|\zeta_j(\tau)\| = \mathsf{y}$ and $\|\zeta_j(t)\| < \mathsf{y}$, for all $t < \tau_j$,
 $j\in \mathcal{Z}_\infty$. 
Fix now constants $\bar{\zeta}_j \in (|\zeta_j(0)|, 1)$ for $j\in \mathcal{Z}_\infty$. 
According to the aforementioned discussion, there exist time instants $\tau_j$ such that $|\zeta_j(\tau_j)| = \bar{\zeta}_j$ and $|\zeta_j(t)| < \bar{\zeta}_j$ for all $t < \tau_j$, $j\in \mathcal{Z}_\infty$. 
By setting $\tau \coloneqq \max_{j\in\mathcal{Z}_\infty}\{\tau_j\}$ and invoking the boundedness of $\zeta_j(t)$, $j\in\{1,\dots,n\}\backslash \mathcal{Z}_\infty$, we conclude the existence of a positive constant $\bar{\zeta} < 1$ such that $|\zeta_j(t)| \leq \bar{\zeta} < 1$ for all $j\in\{1,\dots,n\}$ and $t\in[0,\tau)$. 
Note that $\bar{\zeta}$ is independent of $t_{\max}$ since $\bar{\zeta}_j$ are chosen arbitrarily in $(|\zeta_j(0)|, 1)$, for $j\in \mathcal{Z}_\infty$. 
We further obtain $|\eta_j(t)| = \left| \tfrac{\zeta_j(t)}{\beta_j(t)} \right| \leq \left| \tfrac{\zeta_j(t)}{\beta_j(0)} \right| = \left|\zeta_j(t) \right| \leq \bar{\zeta} < 1$, for all $t\in[0,\tau)$ and $j\in\{1,\dots,n\}$.

For $t\in(\tau,t_{\max})$, it holds that $\beta_j(t) \geq \underline{\beta} > 1$ (by construction) and
 $|\eta_j(t)| = \left| \tfrac{\zeta_j(t)}{\beta_j(t)} \right| \leq \tfrac{1}{\underline{\beta}}  < 1$, $j\in\{1,\dots,n\}$. Therefore, we conclude the boundedness of $\eta_j(t)$ as 
$| \eta_j(t)| \leq \bar{\eta} \coloneqq \max\left\{\bar{\zeta}, \max_{j\in\{1,\dots,n\}}\left\{ \tfrac{1}{\underline{\beta}} \right\}\right\} < 1$ for all  $j\in\{1,\dots,n\}$ and $t\in J$. Note that $\bar{\eta}$ is strictly less than $1$ regardless of the value of $t_{\max}$.

Next, note from \eqref{eq:eta} that 
$s_k$ $=$  $[\Xi^{-1}(\eta_1),\dots,\Xi^{-1}(\eta_n)]^\top$,
where 
$\eta \coloneqq [\eta_1,\dots,\eta_n]^\top$.
Since $|\eta_j|\leq \bar{\eta} < 1$, for all $j\in\{1,\dots,n\}$, $t\in J$, it holds that $\|s_k(t)\| \leq \bar{s}_k$, for all $t\in J$ and a constant $\bar{s}_k$.
Therefore, in view of \eqref{eq:s_i} and Lemma \ref{lemma:s bounds},
we conclude 
the boundedness of $e(t)$ as $\|e(t) \| \leq \bar{E} \coloneqq \bar{e}_1 + \bar{e}_2 \bar{s}_k$ for positive constants $\bar{e}_1$, $\bar{e}_2$, and hence of $x(t)$ as $\|x(t)\| \leq \bar{X} \coloneqq \bar{E} + \sup_{t\geq 0}\{\|\bar{x}_\textup{d}(t)\|\}$ for all $t\in J$.
Furthermore, the boundedness of $x(t)$ and Assumption \ref{ass:internal dynamics} imply the existence of a positive finite constant  $\bar{z}$ such that $z(t) \in \Omega_z \coloneqq \{ z \in \mathbb{R}^{n_z} : \|z\| \leq \bar{z} \}$ for all $t\in J$. Therefore, since $F(x,z,t)$ and $G(x,z,t)$ are uniformly bounded in $t$ 
in view of Assumption \ref{ass:f(x,t) cont + bounded}, we conclude that 
\begin{subequations} \label{eq:F G Lipschitz}
\begin{align} 
&\hspace{-3mm} \|F(x,z,t) - F(\bar{x}_\textup{d},z,t)\|  \leq L_F \|x - \bar{x}_\textup{d}\| = L_F \|e \| \label{eq:F lipschitz} \\  
&\hspace{-3mm}  \|G(x,z,t) - G(\bar{x}_\textup{d},z,t)\|  \leq L_G \|x - \bar{x}_\textup{d}\| = L_G \|e \| \label{eq:G lipschitz}
\end{align}
\end{subequations}
where $L_F$ and $L_G$ are the Lipschitz constant of $F$  and $G$ in $\{x\in\mathbb{R}^{kn}:\|x\|\leq \bar{X}\}  \times \Omega_z 
\times \mathbb{R}_{\geq 0}$.  {Note that $x(t)$ and $z(t)$ are confined by constants $\bar{X}$ and $\bar{z}$ for $t\in J$ \textit{regardless of the initial condition} $x(0)$. More specifically, for any $x(0) \in \mathbb{R}^{kn}$, it is possible to find a $\bar{\zeta} = \bar{\zeta}(x(0)) < 1$ that satisfies $|\zeta_j(t)| \leq \bar{\zeta}$ for all $j\in\{1,\dots,n\}$  and $t\in J$; $\bar{\zeta}$ dictates then $\bar{X}$ and $\bar{z}$ and consequently $L_F$ and $L_G$ in \eqref{eq:F G Lipschitz}, which holds for any initial condition, resembling the global nature of the control algorithm.  }

Consider next the vector $\tilde{s}_{k-1} \coloneqq [s_1^\top,\dots,{s}^\top_{k-1}]^\top$, whose differentiation, in view of \eqref{eq:s_i+1}, yields 
\begin{align} \label{eq:sk-1 dot}
\dot{\tilde{s}}_{k-1} = -  B  \tilde{s}_{k-1} +  C(s_k)
\end{align}
where $C(s_k) \coloneqq  \begin{bmatrix}
0^\top, \dots, 0^\top, s_k^\top
\end{bmatrix}^\top$, and $B = [B_{i,j}] \in \mathbb{R}^{(k-1)n \times (k-1)n}$, with $B_{i,i} = \lambda$ for $i\in\{1,\dots,(k-1)n\}$, $B_{i,i+1} = -1$ for $i\in\{1,\dots,(k-1)n-1\}$, and $B_{ij} = 0$ otherwise; $B$
is a Z-matrix, whose eigenvalues are equal to $\lambda$, rendering it a non-singular M-matrix\footnote{A nonsingular M-matrix is a matrix whose off-diagonal entries are less than or equal to zero and whose principal minors are positive.}. 
Therefore, according to \cite[pp. 168]{qu2009cooperative},
there exists a positive definite diagonal matrix $P \coloneqq \textup{diag}\{ [ p_{i,j} ]\} \in \mathbb{R}^{(k-1)n \times (k-1)n}$ such that $Q \coloneqq P B + B^\top P$ is positive definite.
Consider now the positive definite 
 function 
\begin{align} \label{eq:V_1}
V_1 = V_{1,1} + V_{1,2}  \coloneqq \alpha_0 \tilde{s}_{k-1}^\top P \tilde{s}_{k-1} +  \frac{1}{2}\|\chi\|^2
\end{align}
where 
$\alpha_0$ is a positive constant to be specified later.
Differentiation of $V_{1,1}$ and use of \eqref{eq:sk-1 dot} leads to 
\begin{align*}
\dot{V}_{1,1} =& - \alpha_0  \tilde{s}_{k-1}^\top Q \tilde{s}_{k-1} + 
  2 \alpha_0  \bar{p}  s_{k-1}^\top s_k \\
\leq & - \alpha_0 \underline{Q} \|\tilde{s}_{k-1}\|^2 +   2 \alpha_0 \bar{p}  s_{k-1}^\top s_k 
\end{align*}
for all $t\in J$, where
$\bar{p} \coloneqq p_{(k-1)n,(k-1)n}$ is the bottom right element of $P$ and $\underline{Q} \coloneqq \lambda_{\min}(Q) > 0$ is the minimum eigenvalue of $Q$.
By further completing the squares for the second term, we obtain
\begin{align} \label{eq:V_11 dot}
\dot{V}_{1,1} \leq  - \alpha_0 \tilde{\alpha}_1 \|\tilde{s}_{k-1}\|^2  + \frac{ \alpha_0  \bar{p}}{\alpha_1} \|s_{k}\|^2
\end{align}
for all $t\in J$, where $\tilde{\alpha}_1 \coloneqq  \underline{Q} - \alpha_1  \bar{p}$ and $\alpha_1$ is a positive constant to be chosen later.
Next, differentiation of $V_{1,2}$ along the solution of the closed-loop system leads to 
\begin{align*}
\dot{V}_{1,2} =&  \chi^\top R_T \big( \beta R_\Xi ( F + G u  + \Phi) + \dot{\beta}\eta   \big)
\end{align*}
for all $t\in J$, where $\Phi(x,t) \coloneqq \sum_{\ell=1}^{k-1} \left( \begin{psmallmatrix}
k- 1\\
\ell
\end{psmallmatrix} \right)
\lambda^\ell e_{k-\ell+1} - {x_\textup{d}^{(k)}}$.
Next, note that $\Xi'(\ast) \leq \frac{1}{\sqrt{\kappa}}$. 
Additionally, it can be concluded that 
$|\eta_j | = |\Xi(s_{k_j})| = \frac{|s_{k_j}|}{\sqrt{s_{k_j}^2 + \sqrt{\kappa}}} \leq  \frac{1}{\sqrt{\kappa}}|s_{k_j}|$, for all $j\in\{1,\dots,n\}$, leading to $\|\eta\| \leq \frac{1}{\sqrt{\kappa}} \|s_k\|$. 
 By further adding and subtracting $ \chi^\top R_T \beta R_\Xi F(\bar{x}_\textup{d},z,t)$ and $\chi^\top R_T \beta R_\Xi F_{\textup{d}}$ and invoking \eqref{eq:F lipschitz} and the boundedness of $\beta(t)$ and $\dot{\beta}(t)$, 
$\dot{V}_{1,2}$ becomes
\begin{align*}
\dot{V}_{1,2} \leq &  \chi^\top R_T \beta R_\Xi G u   +  
L_1 \|R_T \chi\| \sum_{\ell = 1}^{k} \|e_{\ell}\| \\
 &+ 
 L_2 \| R_T \chi \| \| s_k \|  +  \chi^\top R_T \beta R_\Xi F_\textup{d} \\
 &+ {\chi^\top R_T \beta R_\Xi(F(\bar{x}_\textup{d},z,t) - F_\textup{d} - x_\textup{d}^{(k)})} 
\end{align*}
for all $t\in J$, where $L_1 \coloneqq \sup_{t\geq 0}\|{\beta}(t)\|(L_F+ \max_{\ell\in\{1,\dots,k-1\}}\{ \begin{psmallmatrix}
k- 1\\
\ell
\end{psmallmatrix} \lambda^\ell  \})$ and $L_2 \coloneqq  \frac{1}{\sqrt{\kappa}} \sup_{t\geq 0}\|\dot{\beta}(t)\|$. 
Next, we show  that  $\sum_{\ell = 1}^{k} \|e_{\ell}\| $ can be upper bounded by a linear combination of   $\|s_\ell\|$, $\ell \in \{1,\dots,k\}$. Indeed, note first from \eqref{eq:s_i} that $e_i = s_i  - \sum_{\ell=1}^{i-1} \left( \begin{psmallmatrix}
i- 2\\
\ell
\end{psmallmatrix} \right)
\lambda^\ell e_{i-\ell}$
for all $i\in\{1,\dots,k\}$,
as well as 
$e_2 = s_2 - \lambda e_1 = s_2 - \lambda s_1$ (by setting $i=2$). 
Therefore, by applying the aforementioned relation in a recursive manner for $i\in\{1,\dots,k\}$, we conclude that each $e_i$ can be expressed as a linear combination of all $s_1$, $s_2$, $\dots$, $s_i$. 
Hence, by further using the boundedness of $\beta(t)$, $\Xi'(\ast) \leq \frac{1}{\sqrt{\kappa}}$, and \eqref{eq:G lipschitz} and completing the squares, $\dot{V}_{1,2}$ becomes 
\begin{align*}
\dot{V}_{1,2} \leq &  \chi^\top R_T \beta R_\Xi G u   +  
\left( \frac{L_3}{\alpha_2} + { \frac{L_4}{\alpha_3}} \right)  \|R_T \chi\|^2 \\
& +  \alpha_2 L_3 \sum_{\ell = 1}^{k} \|s_{\ell}\|^2 
+  \chi^\top R_T \beta R_\Xi F_\textup{d} + { \alpha_3 L_4 \bar{F}_\textup{d}(z,t)^2} 
\end{align*}
for all $t\in J$, positive constants $L_3$, $\alpha_2$, {$\alpha_3$, and $\bar{F}_\textup{d}(z,t) \coloneqq F(\bar{x}_\textup{d}(t),z,t) - F_\textup{d} - x_\textup{d}(t)^{(k)}$, 
$L_4 \coloneqq \frac{1}{\sqrt{\kappa}}\sup_{t\geq 0}\|\beta(t)\|$}.  We now show that $\|s_k\| $ is upper bounded by $\|R_T \chi\|$. 
Indeed, note from \eqref{eq:eta} and \eqref{eq:inverses and derivatives} that 
\begin{align*}
|s_{k_j}| = \left| \frac{\zeta_j \sqrt{\kappa}  }{ \sqrt{\beta_j^2 - \zeta_j^2} } \right| \leq 
\left| \frac{\zeta_j \sqrt{\kappa} }{ \sqrt{1 - \zeta_j^2} } \right| =
\left|\chi_j \sqrt{\kappa(1 - \zeta_j^2)} \right|
\end{align*}
for all $t\in J$, where we use that $\beta_j(t) \geq \beta_j(0) =1$, for all $t\geq 0$, $j\in\{1,\dots,n\}$. 
Note next from \eqref{eq:inverses and derivatives} that $T'(\zeta_j) = \frac{1 + \zeta_j^2}{(1-\zeta_j^2)^2} \geq 1$ and hence  $|s_{k_j}| \leq | \chi_j T'(\zeta_j)  \sqrt{\kappa (1 - \zeta_j^2)}  |$ 
for all $t\in J$ and $j\in\{1,\dots,n\}$. By further noting that $\sqrt{1 - \zeta_j(t)^2} \leq 1$ for all $t\in J$, we obtain that 
$|s_{k_j}| \leq \sqrt{\kappa} \left|\chi_j T'(\zeta_j)\right|$, $j\in\{1,\dots,n\}$, leading to 
\begin{align} \label{eq:s_k less than R_T chi}
\|s_k\| \leq  \sqrt{\kappa}\|R_T \chi\|
\end{align}
for all $t\in J$.
Therefore, by further using
$\| \tilde{s}_{k-1}\|^2 = \sum_{\ell=1}^{k-1}\|s_\ell\|^2$, 
 $\dot{V}_{1,2}$ becomes
\begin{align*}
\dot{V}_{1,2} \leq &  \chi^\top R_T \beta R_\Xi G u   + \alpha_2 L_3  \| \tilde{s}_{k-1}\|^2
 + { \alpha_3 L_4 \bar{F}_\textup{d}^2} \\ & \hspace{-8mm}  + 
\left(\frac{L_3}{\alpha_2} 
+ \alpha_2  L_3 \sqrt{\kappa} + { \frac{L_4}{\alpha_3}}  \right) \|R_T \chi\|^2 +  \chi^\top R_T \beta R_\Xi F_\textup{d} 
\end{align*}
for all $t\in J$, and after substituting the control law \eqref{eq:control law}  
{
\begin{align} \label{eq:V12 dot breaking point}
\dot{V}_{1,2} \leq & 
 -\underline{\lambda}_G (\mu_g + \hat{d}_1 + \|\hat{d}_2\|^2) \| R_\Xi \beta R_T \chi \|^2  \notag \\
 & +  \alpha_2 L_3  \| \tilde{s}_{k-1}\|^2   + 
L_5 \|R_T \chi\|^2 
 -  \chi^\top R_T \beta R_\Xi G \hat{d}_2 \notag \\
&   +  \chi^\top R_T \beta R_\Xi F_\textup{d} + \alpha_3 L_4 \bar{F}_\textup{d}^2,
\end{align}
for all $t\in J$, 
where $\underline{\lambda}_G$ is the minimum eigenvalue of $\frac{1}{2}(G + G^\top)$ and
$L_5 \coloneqq \left(\tfrac{L_3}{\alpha_2} 
+ L_3 \alpha_2\sqrt{\kappa} + \tfrac{L_4}{\alpha_3} \right)$. }

We now define and analyze $T_V \coloneqq  -  \chi^\top R_T \beta R_\Xi G \hat{d}_2 +  \chi^\top R_T \beta R_\Xi F_\textup{d}$ of \eqref{eq:V12 dot breaking point}. 
As mentioned previously, the aim of the integrator $\hat{d}_2$ is to compensate for the constant term $G_\textup{d}^{-1} F_\textup{d}$ from Assumption \ref{ass:F = 0}.
By writing $F_\textup{d} = G_\textup{d} G_\textup{d}^{-1} F_\textup{d}$, {adding and subtracting $\chi^\top R_T \beta R_\Xi G(\bar{x}_\textup{d},z,t) \hat{d}_2$ and $\chi^\top R_T \beta R_\Xi G_\textup{d} \hat{d}_2$, and using \eqref{eq:F G Lipschitz}, $T_V$ becomes}
\begin{align*}
T_V 
=& - \chi^\top R_T \beta R_\Xi \bigg( G_\textup{d}( \hat{d}_2 - G_\textup{d}^{-1} F_\textup{d}) \notag \\
& {- (G(\bar{x}_\textup{d},z,t)- G)\hat{d}_2 - (G_\textup{d} - G(\bar{x}_\textup{d},z,t))\hat{d}_2 \bigg) } \notag \\ 
 \leq &  - \chi^\top R_T \beta R_\Xi G_\textup{d}( \hat{d}_2 - G_\textup{d}^{-1} F_\textup{d}) \notag \\
 &{ +  L_6 \| R_T \chi \| \| e \| \| \hat{d}_2 \| + L_4 \bar{G}_\textup{d}(z,t)  \|R_T \chi\|\|\hat{d}_2\|}
\end{align*}
for $t\in J$, 
where {$\bar{G}_\textup{d}(z,t)  \coloneqq G_\textup{d} - G(\bar{x}_\textup{d}(t),z,t)$} and 
$L_6 \coloneqq L_GL_4 = \frac{L_G}{\sqrt{\kappa}} \sup_{t \geq 0} \| \beta(t) \|$. 
As proven before, $\sum_{\ell =1}^k \|e_\ell\|$  can be upper bounded by a linear combination of $\|s_\ell\|$, $\ell \in \{1,\dots,k\}$. 
Therefore, in view  $\|e\| \leq \sum_{\ell =1}^k \|e_\ell\|$, we obtain
\begin{align*}
 L_6 \| R_T \chi \| \| e \| \| \hat{d}_2 \| & \leq  L_7 \| R_T \chi \|  \sum_{\ell =1}^k \|s_\ell\|   \| \hat{d}_2 \| \\
&\hspace{-20mm} =  L_7 \| R_T \chi  \|  \sum_{\ell =1}^{k-1} \|s_\ell\|   \| \hat{d}_2  \| + 
 L_7 \| R_T \chi \|  \|s_k\|  \| \hat{d}_2  \|
\end{align*}
for $t\in J$ and a positive constant $L_7$. 
By further completing the squares and using \eqref{eq:s_k less than R_T chi} and $\|\tilde{s}_{k-1}\|^2 = \sum_{\ell =1}^{k-1} \|s_\ell\|^2 $,
we obtain 
\begin{align*}
 L_6 \| R_T \chi \| \| e \| \| \hat{d}_2 \| & \leq 
 L_7\alpha_4 \|\tilde{s}_{k-1} \|^2 +
\frac{ L_7}{\alpha_4 } \|R_T \chi\|^2 \|\hat{d}_2\|^2 
\\ 
&\hspace{-10mm} + \sqrt{\kappa} L_7 \alpha_5 \|R_T \chi\|^2 \|\hat{d}_2\|^2 
+ \sqrt{\kappa}\frac{ L_7}{\alpha_5} \|R_T \chi\|^2 \\ 
 {L_4 \bar{G}_\textup{d} \|R_T \chi\|\|\hat{d}_2\| } &  {\leq  \frac{L_4}{\alpha_6} \|R_T \chi\|^2 \|\hat{d}_2\|^2 + \alpha_6 L_4\bar{G}_\textup{d}}^2
\end{align*}
for positive constants $\alpha_4$, $\alpha_5$, $\alpha_6$ to be defined later. 
Therefore,  
\eqref{eq:V12 dot breaking point} becomes
{
\begin{align}
\dot{V}_{1,2} \leq & - \underline{\lambda}_G(\mu_g + \hat{d}_1 + \|\hat{d}_2\|^2)\|R_\Xi \beta R_T \chi \|^2 \notag \\
& + \left( L_5 + \sqrt{\kappa}\frac{L_7}{\alpha_5}  \right) \|R_T \chi\|^2  \notag
\\
&   + \left( \alpha_2 L_3 +  \alpha_4  L_7  \right)\| \tilde{s}_{k-1}\|^2 \notag\\
& - \chi^\top R_T \beta R_\Xi G_\textup{d}( \hat{d}_2 - G_\textup{d}^{-1} F_\textup{d})  \notag \\
& +  \left(\frac{L_7}{\alpha_4} + L_7\sqrt{\kappa} \alpha_5 + \frac{L_4}{\alpha_6} \right) \|\hat{d}_2\|^2 \|R_T \chi\|^2 \notag \\
 &+ \alpha_3 L_4 \bar{F}_\textup{d}^2 +\alpha_6 L_4 \bar{G}_\textup{d}^2
 \label{eq:V_12 dot final} 
\end{align}
By further 
using the facts that $\|s_k(t)\| \leq \bar{s}_k$ for all $t\in J$, $\beta_j(t) \geq \beta_j(0) =1$, for all $t\geq 0$ and $j\in\{1,\dots,n\}$, and
$R_\Xi = \textup{diag}\left\{ \left[\tfrac{\kappa}{(s_{k_j}^2 + \kappa)\sqrt{s_{k_j}^2 + \kappa}} \right]_{j\in\{1,\dots,n\}} \right\}$, we conclude that $- \underline{\lambda}_G \|R_\Xi \beta R_T \chi \|^2 \leq - \gamma_\Xi \|R_T \chi \|^2$, where $\gamma_\Xi \coloneqq \frac{\underline{\lambda}_G \kappa}{(\bar{s}_k^2 + \kappa)\sqrt{\bar{s}_k^2 + \kappa}}$.
Hence, using \eqref{eq:V_11 dot} and \eqref{eq:s_k less than R_T chi}, we obtain
\begin{align}
\dot{V}_1 \leq & - \gamma_\Xi \mu_g \| R_T \chi \|^2 -  \gamma_\Xi \hat{d}_1 \| R_T \chi \|^2 - L_8\|\hat{d}_2\|^2 \| R_T \chi \|^2
  \notag \\
&  - L_9 \| \tilde{s}_{k-1} \|^2 
- \chi^\top R_T \beta R_\Xi G_\textup{d}( \hat{d}_2 - G_\textup{d}^{-1} F_\textup{d})  \notag \\
 &+ L_{10} \|R_T \chi \|^2+ \alpha_3 L_4 \bar{F}_\textup{d}^2 +\alpha_6 L_4 \bar{G}_\textup{d}^2  \label{eq:V_1_dot}
\end{align} }
where  $L_8 \coloneqq  \gamma_\Xi - (\frac{L_7}{\alpha_3} + \alpha_5 L_4\sqrt{\kappa} + \frac{L_4}{\alpha_6})$, $L_{9} \coloneqq  \alpha_0 \tilde{\alpha}_1 - ( \alpha_2 L_3 +  \alpha_4 L_7)$, and $L_{10} \coloneqq  \kappa \frac{\alpha_0 \bar{p}}{\alpha_1} + ( L_5 + \sqrt{\kappa}\frac{ L_7}{\alpha_5} )$.
By choosing $\alpha_3$ and $\alpha_6$ large enough and $\alpha_5$ small enough, we guarantee that $L_8> 0$. 
Similarly, by choosing $\alpha_1$ small enough such that $\tilde{\alpha}_1 = \underline{Q} - \alpha_1 \bar{p} > 0$ and 
 $\alpha_0$ large enough, we guarantee that $L_9 > 0$. Note that $L_3$, $L_4$, and $L_7$ are independent of $\alpha_i$, $i\in\{1,\dots,6\}$, rendering the aforementioned selections feasible. 
 Note also that $L_{10}$ is positive by construction.

Finally, let the errors $\tilde{d}_1 \coloneqq \hat{d}_1 - \frac{L_{10}}{\gamma_\Xi}$, $\tilde{d}_2 \coloneqq \hat{d}_2 - G_\textup{d}^{-1} F_\textup{d}$, and consider the candidate Lyapunov function
\begin{align} \label{eq:V_2}
V_2 \coloneqq V_1 + \frac{\gamma_\Xi}{2\mu_{d_1}}\widetilde{d}_1^2 + \frac{1}{2\mu_{d_2}} \tilde{d}_2^\top G_\textup{d}^\top \tilde{d}_2. 
\end{align}
By differentiating now $V_2$ and using \eqref{eq:V_1_dot}, \eqref{eq:adaptation law}, and \eqref{eq:adaptation law 2}, 
we obtain 
{
\begin{align} 
\dot{V}_2 
 \leq &  -  \gamma_\Xi \mu_g \| R_T \chi \|^2   + \alpha_3 L_4 \bar{F}_\textup{d}^2 +\alpha_6 L_4 \bar{G}_\textup{d}^2 
\end{align}
which, in view of Assumption \eqref{ass:F = 0}, leads to 
\begin{align} \label{eq:V_2 integral final}
V(t) \leq V(0) - \gamma_\Xi \mu_g \int_0^t \|R_T(\zeta(\tau)) \chi(\tau)\|^2 \textup{d}\tau 
+ \bar{\alpha}\bar{W}
\end{align}
for all $t\in J$ and a positive constant $\bar{W}$, where $\bar{\alpha} \coloneqq L_4\max\{\alpha_3, \alpha_6\}$. 
Therefore, we conclude that $V_2(t) \leq \bar{V} \coloneqq V_2(0) + \bar{\alpha}\bar{W}$ and the existence of positive constants $\bar{s}$, $\bar{\chi}$, 
$\bar{d}_1$, $\bar{d}_2$ such that 
$\|s_i(t)\| \leq \bar{s}$, $i\in\{1,\dots,k-1\}$, $\|\chi(t)\| \leq \bar{\chi}$, $\tilde{d}_1(t) \leq \bar{d}_1$, and 
$\|\tilde{d}_2(t)\| \leq \bar{d}_2$, which implies the boundedness of $\hat{d}_1(t)$ and $\hat{d}_2(t)$ for all $t\in J$.  
Therefore, by using the bounds on $s_i(t)$ and applying recursively \eqref{eq:s_i}, we conclude the boundedness of $e_i(t)$, and hence of $x_i(t)$ for all $i\in\{1,\dots,k-1\}$ and $t\in J$. 
Next, by using the inverse of $T(\ast)$ in \eqref{eq:T transf} and invoking its increasing property, we conclude that 
$| \zeta_j(t) | \leq \bar{\zeta} \coloneqq \frac{\sqrt{1+4\bar{\chi}^2}-1}{2\bar{\chi}} < 1$,  for all $t\in J$ and $j\in\{1,\dots,k\}$. 
Consequently, $|\eta_j(t)| \leq \tfrac{\bar{\zeta}}{\beta_j(t)} \leq \bar{\zeta} < 1$ and $|s_{k_j}(t)| \leq \frac{\bar{\zeta}\sqrt{\kappa}}{\sqrt{1-\bar{\zeta}^2}}$,  for all $t\in J$ and $j\in\{1,\dots,k\}$ and we further conclude the boundedness of the control input $u(t)$ and
$\dot{\hat{d}}_1(t)$, $\dot{\hat{d}}_2(t)$ for all $t\in J$. Finally, using \eqref{eq:s_i} for $i=k$ leads to the boundedness of $e(t)$, and hence of $x(t)$, for all $t\in J$. 
Therefore, we conclude that $\bar{x}(t)$ remains in a compact set $\widetilde{\Omega} \subset \bar{\Omega}$ for all $t\in J$, contradicting $\lim_{t\to t_{\max}}\|\bar{x}\| = \infty$ for 
a finite $ t_{\max} < \infty$. Consequently, $t_{\max} = \infty$ and $J = [0,\infty)$.  
Furthermore, by using the inverse of $\Xi(\ast)$ and of $\Psi(\ast)$ and \eqref{eq:beta},
we conclude the validity of \eqref{eq:s ineq} for all $t\geq 0$.
By further using $\bar{\zeta} < 1$ and \eqref{eq:beta} and \eqref{eq:inverses and derivatives}, we obtain $|s_{k_j}(t)| \leq \frac{\sqrt{\kappa}}{\sqrt{\beta_j(t)^2-1}}$ and $-\phi_j(t)  <  
s_{k_j}(t) < \phi_j(t)$ 
for $j\in\{1,\dots,n\}$ and $t > 0$, establishing the funnel specification. By applying Lemma \ref{lemma:s bounds}, we finally conclude the bounds on $e_{i_j}(t)$, for all $j\in\{1,\dots,n\}$, $i\in\{1,\dots,k-1\}$, and $t > 0$.}

\noindent \textbf{3) Asymptotic stabilization $\lim_{t\to\infty} e(t) \to 0$.} \\
We conclude from the aforementioned discussion and \eqref{eq:V_2 integral final} that $\hat{d}_1(t) = \mu_{d_1}\int_0^t\|R_T(\zeta(\tau))\chi(\tau)\|^2\textup{d}\tau$ is bounded and square integrable, for all $t\in J$.
Clearly, $\hat{d}_1(t)$ is non-decreasing, implying that it converges to a finite limit as $t\to\infty$. 
Additionally, by differentiating $\|R_T\chi\|^2$ and using the closed-loop system equations and the proven boundedness of the closed-loop signals, we conclude the boundedness of $\tfrac{\textup{d}}{\textup{d}t}\|R_T\chi\|^2$, for all $t\in J$, leading to the uniform continuity of $\|R_T\chi\|^2$. Consequently, by invoking Barbalat's Lemma \cite[Lemma 8.2]{Khalil}, we conclude that 
$\lim_{t\to\infty} R_T(\zeta(t))\chi(t) = 0$. By further employing the positive lower bound of $R_T(\cdot)$ (by construction), we conclude that $\lim_{t\to\infty} \chi(t) = 0$ and hence $\lim_{t\to\infty}s_i(t) = 0$ for all $i\in\{1,\dots,k-1\}$. 
Since $T(\ast)$ is increasing for $\ast \in (-1,1)$ and $T(0) = 0$, it holds that $\lim_{t\to\infty}\zeta_j(t) = 0$, and since $\beta_j(t) \geq 1$, for all $j\in\{1,\dots,n\}$,  $\lim_{t\to\infty}\eta_j(t) = 0$, $j\in\{1,\dots,n\}$.
By inverting \eqref{eq:eta} and exploiting the monotonically increasing property of $\Xi$ and $\Xi^{-1}$, we conclude that $\lim_{t\to \infty} s_k(t) = 0$.
Finally, since $s_k$ represents the a stable filter with input $s_k$ and output $e_1$, we conclude that $\lim_{t\to \infty} e(t) = 0$, concluding the proof.
\end{proof}

{ \section{Input-constrained Barrier Integral Control} \label{sec:main results input sat}}

{As standard funnel-type controllers, the proposed BRIC algorithm features a high-gain control scheme that can result in a high transient-state peak of the control input, which might not be possible to implement in a real-world scenario. 
For this reason, we now extend the proposed control scheme to account for input saturation, i.e., we accommodate the constraint $|u_j| \leq u_{\textup{sat}}$ for a positive constant $u_{\textup{sat}}$ and $j\in\{1,\dots,n\}$. The dynamics \eqref{eq:dynamics xk} take the form 
\begin{subequations} \label{eq:dynamics input sat}
\begin{align}
\dot{x}_i &= x_{i+1}, \ \ i\in\{1,\dots,k-1\}  \\
\dot{x}_k &= F(x,z,t) + G(x,z,t)\overline{\textup{sat}}_{u_{\textup{sat}}}(u) \\
\dot{z} &= F_z(x,z,t)  
\end{align}
\end{subequations}
where $\overline{\textup{sat}}_{\mathsf{s}}:\mathbb{R}^n \to [-\mathsf{s}, \mathsf{s}]^n$ is the vector form $\overline{\textup{sat}}_\mathsf{s}(\mathsf{x})$ $\coloneqq$ $[\textup{sat}_\mathsf{s}(\mathsf{x}_1)$,$\dots$,$\textup{sat}_\mathsf{s}(\mathsf{x}_n)]^\top$ $\in\mathbb{R}^{kn}$ for a constant $\mathsf{s} > 0$ and a vector $\mathsf{x}\in\mathbb{R}^n$, and 
$\textup{sat}_\mathsf{s}:\mathbb{R} \to [-\mathsf{s}, \mathsf{s}]$ is the saturation function 
\begin{align}
\textup{sat}_\mathsf{s}(\ast) \coloneqq \begin{cases}
\ast & \textup{ if } |\ast| < \mathsf{s} \\
\mathsf{s} \cdot \textup{sign}(\ast) & \textup{ if } |\ast| \geq \mathsf{s}
\end{cases}
\end{align}
Note that the aforementioned transient-state control-input peak is caused by the term $-(\mu_g + \hat{d}_1 + \|\hat{d}_2\|^2)\beta(t) R_\Xi(s_k)R_T(\zeta)\chi$ of the control law \eqref{eq:control law}, which is used to define the transient trajectory of the system; the integrator $\hat{d}_2$ is used to establish the asymptotic behaviour (convergence of the errors to zero). 
Therefore, to deal with input saturation, we combine the reference modification recently proposed in \cite{fotiadis2023input} with an integrator anti-windup scheme. 
We do note that such an anti-windup scheme results in a controller that is not smooth, as in Sec. \ref{sec:main results}, due to the switching off of the integrator for high errors, but it is \textit{continuous}. }

{We describe now the input-constrained BRIC. 
We rewrite first the control law  \eqref{eq:control law} $u(t) = u_P - \hat{d}_2$, with $u_P \coloneqq [u_{P_1},\dots,u_{P_n}]^\top = -(\mu_g + \hat{d}_1 + \|\hat{d}_2\|^2)\beta(t) R_\Xi(s_k)R_T(\zeta)\chi$. We aim to 
apply the  anti-windup scheme to the integrator $\hat{d}_2$ and impose a saturation level $u_{\textup{sat},P}$, to $u_P$, such that $u_{\textup{sat},P} + \|\hat{d}_2(t)\| \leq u_{\textup{sat}}$. For the saturation of $u_{P}$, we define the reference modification system, inspired by \cite{fotiadis2023input}:
\begin{subequations} \label{eq:reference adj}
\begin{align} 
\dot{\sigma}_{i_j} &= -\gamma_i\sigma_{i_j} + \sigma_{(i+1)j},   \sigma_{i_j}(0) = 0, i\in\{1,\dots,k-1\} \notag \\
\dot{\sigma}_{k_j} &= -\gamma_k\sigma_{k_j} - \Delta u_{P_j} \\
\Delta u_{P_j} &\coloneqq {\textup{sat}}_{u_{\textup{sat},P}}(u_{P_j}) - u_{P_j}  \label{eq:Delta u_pj}
\end{align}
\end{subequations}
for positive constants $\gamma_i$, and all $i\in\{1,\dots,k\}$, $j\in\{1,\dots,n\}$.
We further define $\sigma_i \coloneqq [\sigma_{i_1},\dots,\sigma_{i_n}]^\top\in\mathbb{R}^n$ for all $i\in\{1,\dots,k\}$ and $\sigma \coloneqq [\sigma_1^\top,\dots, \sigma_k^\top]^\top \in \mathbb{R}^{kn}$. The modified reference trajectory is then 
$\widetilde{x}_{\textup{d}} \coloneqq x_{\textup{d}} + \sigma_1 \in\mathbb{R}^n$ and the respective error $e \coloneqq [e_1^\top,\dots,e_k^\top] \in \mathbb{R}^{kn}$, with 
$e_i \coloneqq [e_{i_1},\dots,e_{i_n}]^\top \coloneqq x_i - \widetilde{x}_{\textup{d}}^{(i-1)} \in \mathbb{R}^n$ for all $i\in\{1,\dots,k\}$.
We further define 
$\bar{\widetilde{x}}_{\textup{d}} \coloneqq  [\widetilde{x}_{\textup{d}}^\top, \dots, (\widetilde{x}_{\textup{d}}^{(k-1)})^\top ]^\top = \bar{x}_{\textup{d}} + [\sigma_1^\top,\dots,(\sigma_1^{(k-1)})^\top]^\top
$ satisfying $e = x -\bar{\widetilde{x}}_{\textup{d}}$.
Clearly, when $\Delta u_{P_j}=0$, i.e., $u_{j}$ does not exceed $u_{\textup{sat},P}$, then $\sigma_{i_j}(t)$ converges to zero exponentially fast, recovering the original reference $x_\textup{d}(t)$. 
We now define the input-constrained control algorithm as
\begin{subequations} \label{eq:control algorithm input sat}
\begin{align}  
u &= \overline{\textup{sat}}_{u_{\textup{sat},P}}(u_P) - \hat{d}_2 	\label{eq:control law input sat} \\
u_P &\coloneqq -(\mu_g + \hat{d}_1 + \|\hat{d}_2\|^2)\beta(t) R_\Xi(s_k)R_T(\zeta)\chi \label{eq:U_P input sat} \\
\dot{\hat{d}}_2 &= \begin{cases}
\mu_{d_2} \beta(t) R_\Xi(s_k) R_T(\zeta) \chi, & \textup{ if } \|\chi\| \leq \bar{\chi}, \\
0, & \textup{ if } \|\chi\| > \bar{\chi}
\end{cases} \label{eq:adaptation law input sat} 
\end{align}
\end{subequations}
with $\hat{d}_2(0) = 0$, 
where $\bar{\chi}$ is a positive design constant.
The next theorem summarizes the results of this section.}
}

{ 
\begin{theorem} \label{thm:main theorem input sat}
Let the dynamics \eqref{eq:dynamics input sat} under Assumptions \ref{ass:f(x,t) cont + bounded}-\ref{ass:F = 0} and any initial condition $x(0)$. Then, there exists a constant $\bar{u} \geq u_{\textup{sat},P}$ such that, if $u_{\textup{sat}} \geq \bar{u}$, the input-constrained BRIC algorithm \eqref{eq:control algorithm input sat} 
guarantees the existence of  constants $\bar{\zeta}\in(0,1)$, $\lambda_0 \in (0,\lambda)$, $\bar{e}_{i,1}$, $\bar{e}_{i,2}$ such that (i) $|s_{k_j}(t)| < \phi_j(t)$, (ii) $|e_i(t)| < \bar{e}_{i,1}\exp(-\lambda_0 t) + \bar{e}_{i,2} \frac{\bar{\zeta}\sqrt{\kappa}}{\lambda^{k-i}\sqrt{1-\bar{\zeta}^2}} $, for all $t > 0$, $j\in\{1,\dots,n\}$, $i\in\{1,\dots,k-1\}$,  (iii) $\lim_{t\to \infty} e(t) = 0$, and (iv) the boundedness of all closed-loop signals. 
\end{theorem}
}

{
\begin{remark}[\textbf{Relation among $u_\textup{sat}$ and $u_{\textup{sat},P}$}]
In practice, we expect the integrator $\hat{d}_2(t)$ to be active ($\|\chi(t)\| < \bar{\chi}$) when $u_P$ is not saturated, i.e., $|u_{P_j}(t)| < u_{\textup{sat},P}$ for all $j\in\{1,\dots,n\}$. 
In particular, the high control peak that can saturate $u_P$ is expected to occur in the early transient state, common in high-gain funnel controllers \cite{fotiadis2023input,bechlioulis2008robust,berger2021funnel}.
By setting $\bar{\chi} < \|\chi(0)\|$, we can ensure $\hat{d}_2(t) = 0$ in this early transient state. 
Subsequently, when $\chi(t)$ eventually reaches $\|\chi(t)\| < \bar{\chi}$ and $\hat{d}_2(t)$ becomes active, we expect that $|u_{P_j}(t)| < u_{\textup{sat},P}$ for all $j\in\{1,\dots,n\}$. Therefore, $u_{\textup{sat},P}$ can be set equal to $u_{\textup{sat}}$, reducing conservativeness in the choice of $u_{\textup{sat}}$. This is verified by the simulation results, where $u_P(t)$ is initially saturated by  $u_{\textup{sat}}=u_{\textup{sat},P}$, followed by a system trajectory that satisfies $|u_{j}(t)| < u_{\textup{sat}}=u_{\textup{sat},P}$ when $\|\chi(t)\| < \bar{\chi}$, for all $j\in\{1,\dots,n\}$.
\end{remark}
}

\begin{proof}
{ 
Due to \eqref{eq:adaptation law input sat}, we define the solutions of the closed-loop system in the Filippov sense \cite{fischer2013lasalle}.
By following the proof of Theorem \ref{thm:main theorem}, we conclude the existence of at least a local solution $\bar{x}(t) = [\zeta(t)^\top, x(t)^\top, z(t)^\top, \hat{d}_1(t), \hat{d}_2(t)^\top]^\top$ in a maximal interval of existence $J \coloneqq [0,t_{\max})$, with $t_{\max} > 0$, that is continuously differentiable and satisfies $\bar{x}(t) \in \bar{\Omega} \coloneqq (-1,1)^n \times \mathbb{R}^{(k+1)n +n_z+1}$ for almost all (a.a.) $t\in J$. 
Similar to the proof of Theorem \ref{thm:main theorem}, we assume that $t_{\max} < \infty$ (aiming to reach a contradiction), implying that $\lim_{t\to t_{\max}}|\zeta_j(t)| = 1$ for all $j$ in a set $\mathcal{Z}_\infty \subseteq \{1,\dots, n\}$. 
Additionally, by following identical steps with the proof of Theorem \ref{thm:main theorem}, we conclude the existence of a constant $\bar{\eta} < 1$ such that $|\eta_j(t)| \leq \bar{\eta}$ for a.a.  $t \in J$.}



{ 
Without loss of generality, we assume that $\|\chi(0)\| > \bar{\chi}$. 
Since $\bar{x}(0) \in \bar{\Omega}$, it holds that $|u_{P_j}(0)| \leq u_{\textup{sat}} + M$ for a positive constant $M$, $j\in\{1,\dots,n\}$.
Additionally, let a  constant $\bar{z}_0$ satisfying $\bar{z}_0 > \|z(0)\|$  and 
consider a compact set $\mathcal{X} \coloneqq \{ \bar{x} \in \bar{\Omega}: \|z\| \leq \bar{z}_0\}$ such that $\bar{x} \in \mathcal{X}$ implies $|u_{P_j}| \leq u_{\textup{sat}} + M, \ \forall j \in \{1,\dots,n\}$. Note that $\bar{x}(0) \in \mathcal{X}$. We now derive a feasibility condition on $u_{\textup{sat},P}$ to guarantee that $\mathcal{X}$ is positively invariant. 
To this end, note first that the continuity of the solution implies the existence of a positive time instant $t_{\max}' \in (0,t_{\max})$ such that 
$\bar{x}(t) \in \mathcal{X}$ and  $|u_{P_j}(t)| \leq u_{\textup{sat},P} + M$, for all $j\in\{1,\dots,n\}$, and a.a. $t\in J_T \coloneqq [0,t_{\max}']$. 
Therefore, in view of \eqref{eq:Delta u_pj}, we  conclude that 
$|\Delta u_{P_j}(t)| \leq M$, for a.a. $t \in J_T$, 
and $j\in\{1,\dots,n\}$. Hence, by using \eqref{eq:reference adj}, we conclude that 
\begin{align} \label{eq:sigma ij bound}
|\sigma_{i_j}(t)| \leq \frac{M}{\Pi_{\ell = i}^k \gamma_{\ell}} 
\end{align}
for  a.a. $t \in J_T$, $j\in\{1,\dots,n\}$, $i\in\{1,\dots,k\}$. 
Moreover, since  $\mathcal{X} \subset \bar{\Omega}$, it holds that 
$|\eta_j(t)| \leq \bar{\bar{\eta}}$, $j\in\{1,\dots,n\}$, for a constant $\bar{\bar{\eta}} \leq \bar{\eta} < 1$ and  a.a. $t\in J_T$. Consequently, it holds that $|s_{k_j}(t)| = | \Xi^{-1}(\eta_j(t)) | =  \left| \tfrac{\eta_j(t)\sqrt{\kappa}}{\sqrt{1-\eta_j(t)^2}} \right| \leq  \tfrac{\bar{\bar{\eta}}\sqrt{\kappa}}{\sqrt{1 - \bar{\bar{\eta}}^2}}$ and $\|s_k(t)\| \leq \sqrt{n}\tfrac{\bar{\bar{\eta}}\sqrt{\kappa}}{\sqrt{1 - \bar{\bar{\eta}}^2}} =: \bar{s}_{k}$,  for $j\in\{1,\dots,n\}$ and a.a. $t\in J_T$. 
By invoking \eqref{eq:s_i} and Lemma \ref{lemma:s bounds}, we conclude the boundedness of $e(t)$ as $\|e(t)\| \leq \bar{E} \coloneqq \bar{e}_1 + \bar{e}_2 \bar{s}_k$ for positive constants $\bar{e}_1$ and $\bar{e}_1$ dependent on $e(0)$, $\lambda$, and $\bar{s}_k$. 
Consequently, and in view of \eqref{eq:reference adj} and \eqref{eq:sigma ij bound}, it holds that $\|x(t)\| \leq \bar{X} \coloneqq \bar{E} + \sup_{t\geq 0}\{\bar{x}_\textup{d}(t)\} + \bar{\gamma} M$ for a.a. $t\in J$, where $\bar{\gamma}$ is a term dependent on $\gamma_i$, $i\in\{1,\dots,k\}$.
Similarly to the proof of Theorem \ref{thm:main theorem},  the boundedness of $x(t)$ and Assumption \ref{ass:internal dynamics} imply the existence of a positive finite constant  $\bar{z}$, dependent on $M$, such that $\|z(t)\| \leq \bar{z}$ for a.a. $t\in J$. 
Consequently, the Lipschitz inequalities \eqref{eq:F G Lipschitz} hold for a.a.  $t\in J_T$ and some constants $L_F$, $L_G$.  }

{ 
We now prove that $\|\chi(t_f)\| = \bar{\chi}$ for a finite constant $t_f < t_{\max}'$. Aiming for a contradiction, assume that $\|\chi(t)\| > \bar{\chi}$  for a.a. $t\in J_T$.
Consider the function 
$V_{s_1} \coloneqq  \frac{1}{2}\|\chi\|^2$. Assume that 
 $|u_{P_{j^\ast}}| \geq u_{\textup{sat},P}$ for at least one $j^\ast \in \{1,\dots,n\}$.
By following similar steps as in the proof of Theorem \ref{thm:main theorem}, adding and subtracting the term 
$\chi^\top R_T\beta R_\Xi G u$, and using the fact that $\hat{d}_2(t) = 0$, for a.a. $t\in J_T$, we can conclude that
\begin{align*}
\dot{V}_{s_1} \leq & - \gamma_\Xi(\mu_g + \hat{d}_1)\| R_T \chi \|^2 + \|R_T\chi\| D_f  \\
 & + \chi^\top R_T \beta R_\Xi\big( (G - I) \Delta u_P + \Lambda(\sigma) \big)  \\
 \end{align*} 
for a.a. $t\in J_T$, where $\Lambda(\sigma) \coloneqq  - \sum_{\ell = 0}^{k-1} \gamma_{k-\ell} \sigma_1^{(\ell)}$, $\Delta u_P \coloneqq [\Delta u_{P_1}^\top,\dots, \Delta u_{P_n}]^\top$,
and $D_f$ is a positive constant satisfying $D_f \geq \| \beta R_\Xi(F + \Phi) + \dot{\beta}\eta \|$ due to the aforementioned boundedness of $x(t)$, $s_k(t)$, $z(t)$ and $e(t)$ for a.a. $t\in J_T$. We further use $\Phi(x,t) \coloneqq \sum_{\ell=1}^{k-1} \left( \begin{psmallmatrix}
k- 1\\
\ell
\end{psmallmatrix} \right)
\lambda^\ell e_{k-\ell+1} - {x_\textup{d}^{(k)}}$ as in the proof of Theorem \ref{thm:main theorem}. 
For $j^\ast$, since $|u_{P_{j^\ast}}(t)| \geq u_{\textup{sat},P}$, 
we obtain that $ |\Delta u_{P_{j^\ast}}(t) | \leq M \leq \frac{M}{u_{\textup{sat},P}} |u_{P_{j^\ast}}(t)|$ leading to
\begin{align} \label{eq: M bound}
\| \Delta u_P(t) \| \leq  \frac{M\sqrt{n}}{u_{\textup{sat},P}} \|u_P(t)\| 
\end{align}
for a.a.  $t \in J_T$.
By using \eqref{eq:reference adj} - \eqref{eq: M bound} and the boundedness of $G(x,z,t)$ due to the boundedness of $x(t), z(t)$ and Assumption \ref{ass:f(x,t) cont + bounded}, we obtain
\begin{align*}
& \chi^\top R_T \beta R_\Xi\big( (G - I ) \Delta u_P + \Lambda(\sigma) \big) \leq & \\ 
& \hspace{25mm} \frac{L_{s_1}}{u_{\textup{sat},P}} \|\beta R_\Xi R_T \chi\| \|u_P\|, 
\end{align*}
where $L_{s_1} \coloneqq M\sqrt{n}(\max_{\|x\|\leq \bar{X}, \|z\|\leq \bar{z}} \{G(x,z,t) + I\}$ $+ \bar{\Gamma})$ and $\bar{\Gamma}$ is a function of the constants $\gamma_i$, $i\in\{1,\dots,k\}$.
By substituting $u_P$, we obtain 
\begin{align} \label{eq:Delta u ineq}
\chi^\top R_T \beta R_\Xi\big( (G - I ) \Delta u_P + \Lambda(\sigma) \big)  \leq & \notag  \\
& \hspace{-25mm} \frac{ L_{s_2} }{u_{\textup{sat},P}} (\mu_g + \hat{d}_1) \| R_T \chi\|^2  
\end{align}
for a.a. $t\in J_T$, where $L_{s_2} =  L_{s_1} \frac{1}{\sqrt{\kappa}}\sup_{t\geq 0}\|\beta(t)\|$. Therefore, $\dot{V}_{s_1}$ becomes
\begin{align*}
\dot{V}_{s_1} \leq & -\left(\gamma_\Xi - \frac{L_{s_2}}{u_{\textup{sat},P}} \right) (\mu_g + \hat{d}_1)\| R_T \chi \|^2 + \|R_T\chi\| D_f 
\end{align*}
for a.a. $t\in J_T$.	
By choosing 
\begin{align*} 
u_{\textup{sat},P} \geq \bar{u}_1 \coloneqq \frac{L_{s_2}}{\gamma_\Xi-\epsilon_{u_1}},
\end{align*}
for a positive constant $\epsilon_{u_1} < \gamma_\Xi$, we obtain 
\begin{align*}
-\gamma_\Xi +  \frac{ L_{s_2} }{u_{\textup{sat},P}} \leq -\gamma_\Xi +  \frac{ L_{s_2}  }{\bar{u}_1} = -\epsilon_{u_1}
\end{align*}
leading to 
\begin{align} \label{eq:Vs1 dot} 
\dot{V}_{s_1}  \leq & - \epsilon_{u_1}(\mu_g + \hat{d}_1)\|R_T \chi\|^2+ \|R_T\chi\| D_f. 
\end{align}
By using \cite[Theorem 4.18]{Khalil}, we conclude the boundedness of $\chi(t)$, and hence of $R_T(\zeta(t))$, as $\|\chi(t)\| \leq \bar{D}_\chi$ and $\|R_T(\zeta(t))\| \leq \bar{D}_T$ for a.a. $t\in J_T$ and positive constants $\bar{D}_\chi$, $\bar{D}_T$, independent of $t_{\max}$. Therefore, by further noticing from \eqref{eq:inverses and derivatives} that $T'(\ast) \geq 1$, $\dot{V}_{s_1}$ becomes
\begin{align*}
\dot{V}_{s_1} \leq - \epsilon_{u_1} (\mu_g + \hat{d}_1)\|\chi\|^2 + \bar{D}_V,
\end{align*}
 for a.a. $t\in J_T$, where $\bar{D}_V \coloneqq \bar{D}_T \bar{D}_\chi D_f$. Since $T'(\ast) \geq 1$ and we have assumed that $\|\chi(t)\| \geq \bar{\chi}$, it holds that $\hat{d}_1(t) \geq t \mu_{d_1} \bar{\chi}^2$, for a.a. $t\in J_T$, and consequently
\begin{align*}
\dot{V}_{s_1} \leq - \epsilon_{u_1}( \mu_g + t \mu_{d_1}\bar{\chi}^2)\bar{\chi}^2 + \bar{D}_V.
\end{align*}
Therefore, $\dot{V}_{s_1}(t) < 0$ for a.a. $t \geq t_n \coloneqq   \frac{ \bar{D}_V - \bar{\chi}^2  \mu_g }{\epsilon_{u_1}\mu_{d_1} \bar{\chi}^4}$. Clearly, since $V_{s_1} = \frac{1}{2}\|\chi\|^2$, there exists a time instant $t_f \in (t_n,t_{\max})$ such that $\|\chi(t_f)\| = \bar{\chi}$, leading to a contradiction.
There further exists an input-saturation level $\bar{u}_2$ such that $u_{\textup{sat},P} \geq \bar{u}_2$ leads to a large enough $t_{\max}'$ yielding $t_f < t_{\max}'$.
A similar analysis can be performed for the case $|u_{P_{j}}(t)| \geq u_{\textup{sat},P}$ for all $j\in\{1,\dots,n\}$.} 

{
We consider next the case $\|\chi\| \leq \bar{\chi}$ and $|u_{P_{j^\ast}}(t)| \geq u_{\textup{sat},P}$ for at least one $j^\ast \in\{1,\dots,n\}$\footnote{{The case $|u_{P_j}(t)| \leq u_{\textup{sat},P}$ for all $j^\ast \in\{1,\dots,n\}$ is covered by Theorem \ref{thm:main theorem}.}}. 
Similar to \eqref{eq:V_2}, consider the Lyapunov function 
\begin{align*}
V_{s_2} \coloneqq V_1 + \frac{\gamma_\Xi - \mu_f}{2\mu_{d_1}}\widetilde{d}_1^2 + \frac{1}{2\mu_{d_2}}\widetilde{d}_2^\top G_\textup{d}^\top \widetilde{d}_2,
\end{align*}
where $\widetilde{d}_1 \coloneqq \hat{d}_1 - \frac{L_{10}}{\gamma_\Xi - \mu_f}$, $\widetilde{d}_2 \coloneqq \hat{d}_2 - G_\textup{d}^{-1}F_\textup{d}$, $\mu_f$ is a positive constant satisfying $\mu_f < L_8 < \gamma_\Xi$, and $\gamma_\Xi$, $L_8$, $L_{10}$ as defined in \eqref{eq:V_1_dot} in the proof of Theorem 1.
By differentiating $V_{s_2}$, adding and subtracting $\chi^\top R_T \beta R_\Xi Gu$, taking into account \eqref{eq:Delta u ineq}, and following the same steps in the proof of Theorem \ref{thm:main theorem}, we obtain
\begin{align*}
\dot{V}_{s_2} \leq & -\left(\mu_f - \frac{L_{s_2}}{u_{\textup{sat},P}}\right)(\mu_f + \hat{d}_1 + \|\hat{d}_2\|^2)\|R_T\chi\|^2 \\
&+ \alpha_3 L_4 \bar{F}_\textup{d}^2 + \alpha_6 L_4 \bar{G}_\textup{d}^2,  
\end{align*}
with $\alpha_3$, $L_4$, $\bar{F}_\textup{d}$, $\alpha_6$, $\bar{G}_\textup{d}$ as defined in the proof of Theorem \ref{thm:main theorem}. Therefore, by choosing 
\begin{align*} 
u_{\textup{sat},P} \geq \bar{u}_3 \coloneqq \frac{L_{s_2}}{\mu_f-\epsilon_{u_2}},
\end{align*}
for a positive constant $\epsilon_{u_2} < \mu_f$, we obtain 
\begin{align*}  
\dot{V}_{s_2}  \leq & - \epsilon_{u_2}(\mu_g + \hat{d}_1 + \|\hat{d}_2\|^2)\|R_T \chi\|^2 \\
&+ \alpha_3 L_4 \bar{F}_\textup{d}^2 + \alpha_6 L_4 \bar{G}_\textup{d}^2. 
\end{align*}
for a.a. $t\in J_T$. Finally, by following similar arguments as in the proof of Theorem \ref{thm:main theorem}, we conclude the extension of $t_{\max}$ to $\infty$, the boundedness of all closed-loop signals, the establishment of the funnel constraints for $s_k(t)$ and $e(t)$, and the convergence of $\chi(t)$ and $e(t)$ to zero. 
Therefore, it holds that $|\zeta_j(t)| < \bar{\zeta}_f$, $|{\chi_j(t)}| \leq \bar{X}_f$, $|z_\ell(t)| < \bar{z}_f$, 
$|\hat{d}_1(t)| \leq \bar{D}_{1_f}$, and $|\hat{d}_{2_j}(t)| \leq \bar{D}_{2_f}$, for a.a. $t\geq 0$, $j\in\{1,\dots,n\}$, $\ell\in\{,\dots,n_z\}$, and positive constants
$\bar{\zeta}_f < 1$, $\bar{X}_f$, $\bar{z}_f$, $\bar{D}_{1_f}$, $\bar{D}_{2_f}$.
Therefore, we conclude that 
$\bar{x}(t) \in \mathcal{X}_f \coloneqq [-\bar{\zeta}_f,\bar{\zeta}_f]^n \times [-\bar{X}_f,\bar{X}_f]^n \times[-\bar{z}_f,\bar{z}_f]^{n_z} \times [-\bar{D}_{1_f},\bar{D}_{1_f}]\times[-\bar{D}_{2_f},\bar{D}_{2_f}]^n$ for a.a. $t\geq 0$. 
In view of the set $\mathcal{X} = \{\bar{x} \in \bar{\Omega}: \|z\| \leq \bar{z}_0$\}, where it holds  $\ |u_{P_J}| \leq u_{\textup{sat},P} + M$, for all  $j \in \{1,\dots,n\}$, defined in the beginning of the proof, we conclude that, by choosing large enough $\bar{z}_0$, there exists a finite $\bar{u}_4 > 0$ such that  $u_{\textup{sat},P} \geq \bar{u}_4$ implies that $\mathcal{X}_f \subseteq \mathcal{X}$.  
Hence, by choosing $u_{\textup{sat},P} \geq \bar{u}_P \coloneqq \max\{\bar{u}_1,\bar{u}_2, \bar{u}_3, \bar{u}_4, \max_{j\in\{1,\dots,n\}} \{|u_{P_j}(0)| - M\}\}$, we conclude that the set $\mathcal{X}$ is positively invariant. 
By further choosing $u_{\textup{sat}} \geq \bar{u} \coloneqq \bar{u}_P + \bar{D}_{2_f}$ to account for the integrator $\hat{d}_2$, we guarantee that $|u_j(t)| \leq u_{\textup{sat}}$ for a.a. $t \geq 0$ and all $j\in\{1,\dots,n\}$, leading to the conclusion of the proof.
}

\end{proof}

{
\begin{remark}[\textbf{Robustness of} $\hat{d}_1$, $\hat{d}_2$]
By inspecting the proof of Theorems \ref{thm:main theorem} and \ref{thm:main theorem input sat}, one can suspect that the \textit{asymptotic} guarantees no longer hold if Assumption \ref{ass:F = 0} does not hold, which can occur in practice.
Nevertheless, in such cases, one can derive a relation similar to \eqref{eq:Vs1 dot}, which shows the boundedness and decrease of $\chi(t)$ as $\hat{d}_1(t)$ increases. 
In fact, one can show that, for any $\epsilon > 0$, there exists a finite time instant $t_\epsilon$ such that $\|\chi(t)\| \leq \epsilon$ for all $t\geq t_\epsilon$, implying asymptotic convergence of $\chi(t)$ to zero, even without the inclusion of $\hat{d}_2(t)$. 
Although this does not prove the boundedness of $\hat{d}_1(t)$, which can diverge as $\lim_{t\to\infty}\hat{d}_1(t) = \infty$, 
the product $\hat{d}_1(t) \chi(t)$ used in the control design is expected to remain bounded due to the convergence of $\chi(t)$. Intuitively, the increase of $\hat{d}_1(t)$ induces a feedback term in $u(t)$ with higher gain, leading to a decrease of $\chi(t)$.  By further using the input-constrained BRIC version, this does not come at a cost of arbitrarily high or unbounded control inputs. 
Nevertheless, it should be noted that the control law in this case uses the product of a term $\hat{d}_1(t)$ diverging to infinity  and a term $\chi(t)$ converging to zero, (similar to \cite{lee2019asymptotic,zhou2023asymptotic}), which can lead to numerical issues.
Additionally, an oscillatory asymptotic behaviour for $\chi(t)$, which is typical for system affected by disturbances, is expected to lead to bounded $\hat{d}_2(t)$.   In practical scenarios with significant disturbances, any potential divergence of  ${\hat{d}}_1(t)$, ${\hat{d}}_2(t)$ can be avoided by using suitable adaptive-control methods, such as $\sigma$-modification \cite{tsakalis1992sigma}, or  
by ``switching off" ${\hat{d}}_1(t)$ and ${\hat{d}}_1(t)$ when they exceed a certain value.
Simulation results verify the aforementioned observations, since they show that BRIC drives the transformed error $\chi(t)$ to zero, retaining the boundedness of $\dot{\hat{d}}_1(t)$ and $\dot{\hat{d}}_2(t)$ for cases where Assumption \ref{ass:F = 0} is not satisfied.
\end{remark}
}

\section{Simulation Results} \label{sec:sims}

To demonstrate the BRIC algorithm, we consider the control problem of two inverted pendulums connected by a spring and a damper \cite{verginis2020asymptotic}. Each pendulum is controlled by a torque input applied by a servomotor at its base. The equations of motion in terms of the angular positions of the pendulums $\theta = [\theta_1, \theta_2]^\top$ and applied torques $u = [u_1, u_2]^\top$ are given, for $i\in\{1,2\}$, by:
\begin{align} \label{eq:pendulums ode}
& J_i \ddot{\theta}_i = r_c \big( g  m_i \sin\theta_i +  (-1)^i 0.5 F_c  \cos(\theta_i - \theta_c) \big) - T_i \notag  \\ 
&\hspace{40mm} + {d_i(t)} + B_{c_i}(q,t)^\top u
\end{align}
where $F_c = k_c (x_c - l_c) + b_c\dot{x}_c$ represents the force applied by the spring and the damper at the connection points; $x_c$ denotes their distance given by 
\begin{align*}
x_c = \sqrt{d_c^2 + d_c r_c (\sin\theta_1 - \sin\theta_2) + \frac{r_c^2}{2}(1 - \cos(\theta_2 - \theta_1))}
\end{align*}
and $\theta_c$ is $\theta_c = \tan^{-1}\left( \frac{ r_c(\cos\theta_2-\cos\theta_1) }{ 2d_c + r_c (\sin\theta_1 - \sin\theta_2) } \right)$.
The terms $T_i$ correspond to friction terms, assumed to follow the LuGre model $T_i = \sigma_0 \tau_i + \sigma_1 \dot{\tau}_i + \sigma_2 \dot{\theta}_i$, with
\begin{align*}
\dot{\tau}_i =& \dot{\theta}_i - \sigma_0 \frac{|\dot{\theta}_i|}{ T_c + (T_s - T_c)\exp( - |\frac{\dot{\theta}_i}{\dot{\theta}_s}|^2 ) } 
\end{align*}
for $i\in\{1,2\}$. {The terms $d_i(t)$ represent vanishing disturbances, chosen as $d_1(t) = 2\exp(-0.01t)\sin(2t+\frac{\pi}{4})$ and $d_2(t) = 2\exp(-0.01t)\cos(2t - \frac{\pi}{6})$. }
 Finally, we set the input matrix as $B_{c_1}$ $=$  $ [\cos\theta_1 + 1.5$, $-\cos\theta_2 \sin\theta_1]^\top$, $B_{c_2}$ $=$ $[-\cos\theta_2 \sin\theta_1$, $\sigma_t(t) \sin\theta_2 \cos\theta_2 + 2]^\top$,
where $\sigma_t$ models a temporary motor failure as $\sigma_t(t) = 0.5$, if $t\in [2,10)$, and $\sigma_t(t) = 1$ otherwise. 
Finally, the parameters $J_i$, $m_i$, $g$, $r_c$, $d_c$, represent moments of inertia, masses, gravity acceleration, pendulum length, and the distance among the pendulum bases, respectively. 

We conduct a comparative simulation study with the funnel-based Prescribed Performance Control (PPC) algorithm \cite{bechlioulis2008robust}. We choose the parameters and initial conditions as $J_1 = 0.5$, $J_2 = 0.625$, $m_1 = 2$, $m_2 = 2.5$, $r_c = 0.5$, $d_c = 0.5$, $l_c = 0.5$, $k_c = 150$, $b_c = 1$, $g = 9.81$, $\sigma_0 = \sigma_1  = 1$, $\sigma_2 = 1$, $\dot{\theta}_s = 0.1$, $T_s = 2$, $T_c = 1$, $\dot{\theta}_1 = \dot{\theta}_2 = 0$,  $\dot{\theta}_1 = -1.6$, $\dot{\theta}_2 = 0.96$ rad.
We choose the reference {trajectory as $x_{\textup{d}_1}(t) = -\frac{\pi}{4} - \frac{\pi}{6} \cos(\frac{3}{2}t)\exp(-0.1t)$, $x_{\textup{d}_2}(t) = \frac{\pi}{4} + \frac{\pi}{6} \cos(t) \exp(-0.1t)$}, 
 and the control gains and parameters as $\phi_1(t) = \phi_2(t) = \frac{1}{t}\exp(-0.5t) + 0.5$, $\kappa = 20$, $\lambda=1$, $\mu_g = 0.1$, $\mu_{d_1} = 10$, $\mu_{d_2} = 20$.
For the PPC algorithm, we choose the funnel as $\rho_1(t) = \rho_2(t) = \|s_2(0)\| \exp(-0.5t) + 0.5$ and the control gain as $k_{ppc}= \mu_g = 0.1$.

\begin{figure}
\centering
\includegraphics[trim={0mm 0 0 0},clip,scale=0.475]{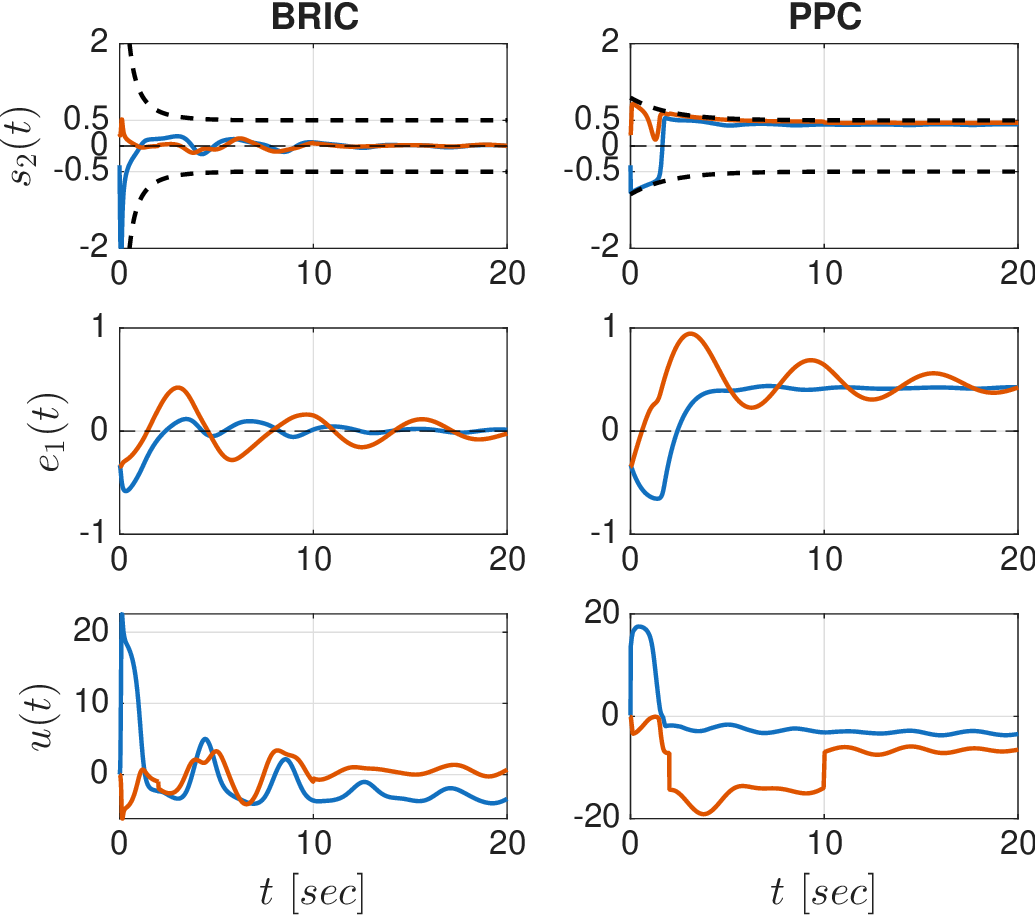}
\caption{The errors $s_2(t)$ (top),  $e_1(t)$ (middle), and control inputs $u(t)$ (bottom) for  BRIC (left) and PPC (right), along with the barrier functions $\phi(t)$ and $\rho(t)$ (top).  }
\label{fig:errors comparison}
\end{figure}

\begin{figure}
\centering
\includegraphics[scale=0.5]{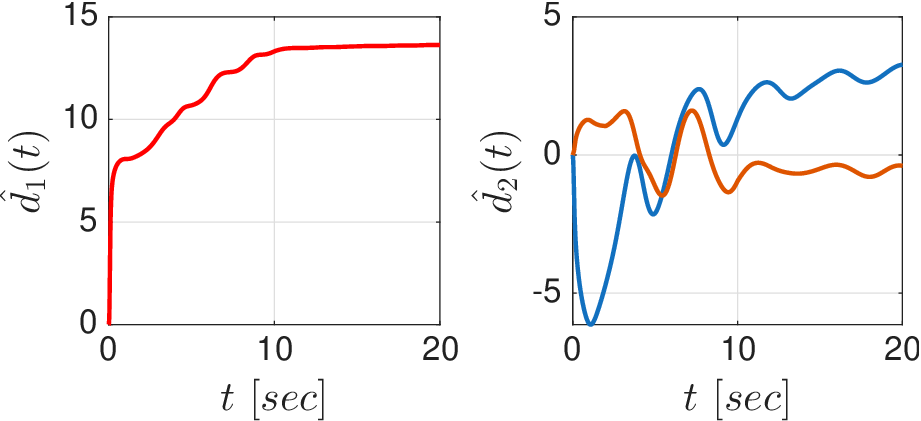}
\caption{The integral signals $\hat{d}_1(t)$ and $\hat{d}_2(t)$.}
\label{fig:dhats}
\end{figure}

The simulation results are shown in Figs. \ref{fig:errors comparison} and \ref{fig:dhats}  for $20$ seconds.
More specifically, Fig. \ref{fig:errors comparison} depicts the evolution of the errors $s_2(t)$, $e_1(t)$ for both BRIC and PPC algorithms, along with the barrier functions  $\pm \phi(t)$ and $\pm \rho(t)$, respectively, as well as the control input $u(t)$.
It is clear that the BRIC algorithm outperforms PPC in terms of steady-state error - since it successfully drives $s_2(t)$ and $e(t)$ to zero. 
Additionally, Fig. \ref{fig:dhats} depicts the evolution of the integral signals $\hat{d}_1(t)$, $\hat{d}_2(t)$ for the BRIC algorithm. Note that $\hat{d}_1(t)$ converges to a constant value due to the convergence of $s_2(t)$ to zero.

\begin{figure}
\centering
\includegraphics[trim={1.6cm 0 0 0}, clip, scale=0.375]{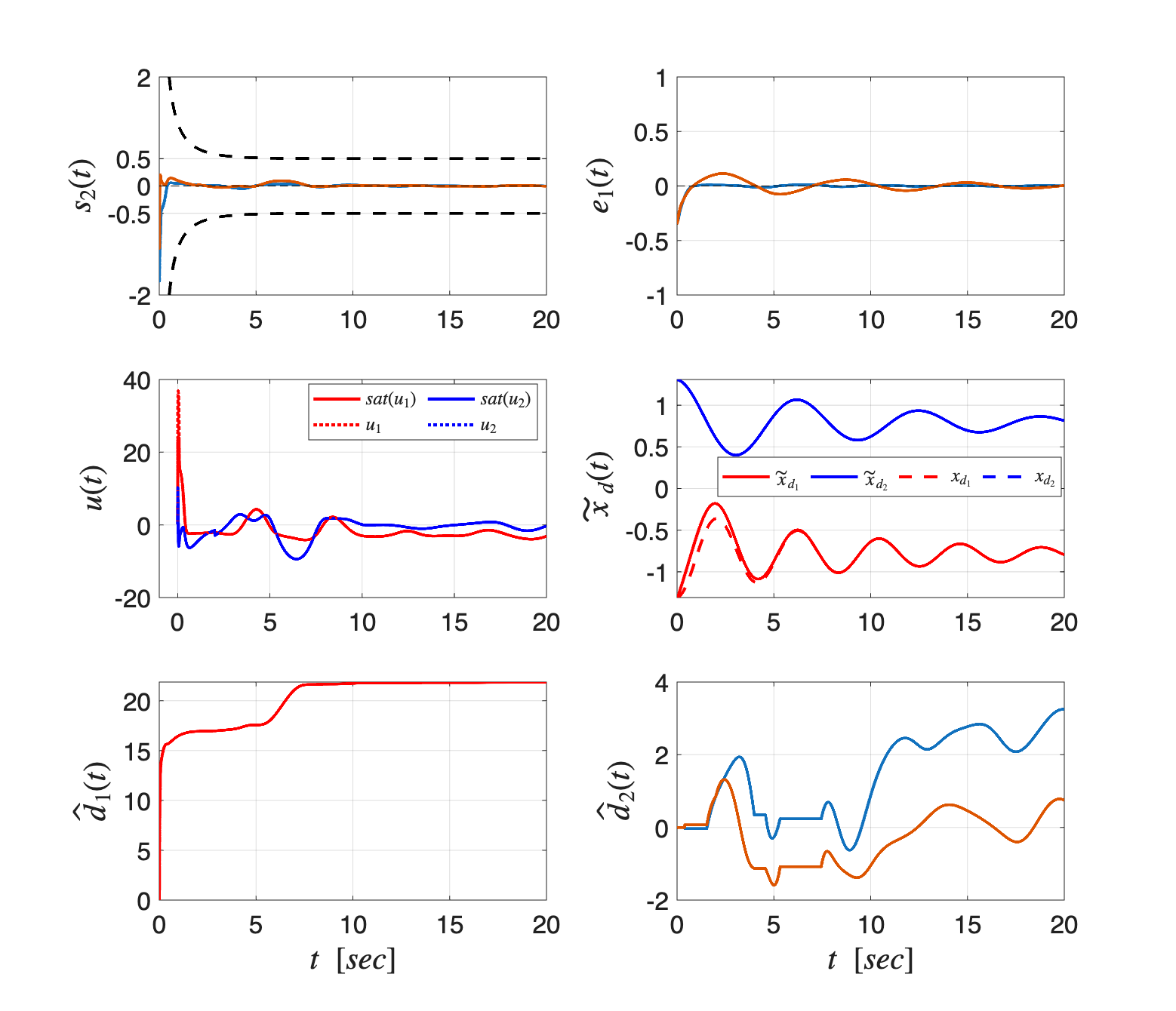}
\caption{The errors $s_2(t)$ (top),  $e_1(t)$ (middle), and control inputs $u(t)$ (bottom) for  BRIC (left) and PPC (right), along with the barrier functions $\phi(t)$ and $\rho(t)$ (top).  }
\label{fig:data_sat}
\end{figure}

{We next evaluate the input-constrained BRIC algorithm. To that end, we set $\kappa = 3$, $\lambda=5$, $\mu_g = 0.1$, $\mu_{d_1} = 10$, $\mu_{d_2} = 20$ in order to create a large initial control-input spike. We further set $u_{\textup{sat}} = u_{\textup{sat},P} = 25$ and $\bar{\chi} = 0.1$. 
The results are given in Fig \ref{fig:data_sat}, which depicts the asymptotic convergence of $s_2(t)$ and $e_1(t)$ to zero and the evolution of $\hat{d}_1(t)$, $\hat{d}_2(t)$. Additionally, it depicts the required and actual control input, $u(t)$ and $\overline{\textup{sat}}(u(t))$ as well as the reference-modified signal $\widetilde{x}_\textup{d}(t) = x_\textup{d}(t) + \sigma_1(t)$.  Despite the input saturation, the reference modification scheme manages to successfully drive the errors to zero. We note that we set $u_{\textup{sat}} = u_{\textup{sat},P}$ since we expect the input-saturation for $u_P$ is expected to occur initially, when $\hat{d}_2(t)$ is still zero, which is verified by the results. }

\begin{figure}
\centering
\includegraphics[scale=0.35]{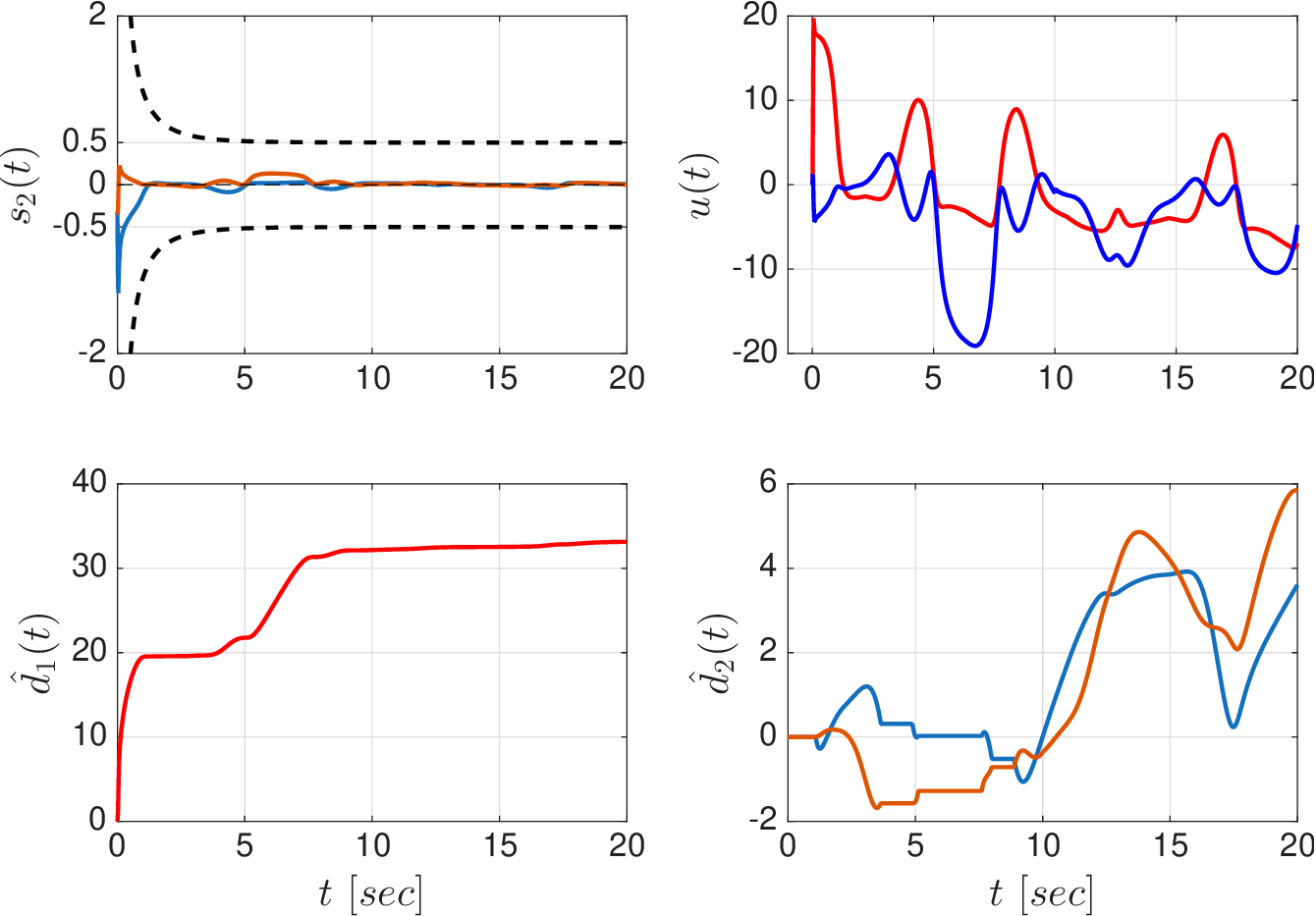}
\caption{The errors $s_2(t)$ (top left), required and saturated control inputs $u(t)$, $\overline{\textup{sat}}(u(t))$ (top right), and adaptation variables $\hat{d}_1(t)$, $\hat{d}_2(t)$ for $q_R > 0$ and references and dynamics that do not satisfy Assumption \ref{ass:F = 0}.} \label{fig:data_sat_dist}
\end{figure}


Finally, we show the robustness of BRIC against disturbances a reference that do not satisfy Assumption \ref{ass:F = 0}, {namely $d_1(t) = 2\sin(2t+\frac{\pi}{4})$ and $d_2(t) = 2\cos(2t - \frac{\pi}{6})$, $x_{\textup{d}_1}(t) = -\frac{\pi}{4} - \frac{\pi}{6} \cos(\frac{3}{2}t)$, and $x_{\textup{d}_2}(t) = \frac{\pi}{4} + \frac{\pi}{6} \cos(t)$. }
 The results are shown in Fig. \ref{fig:data_sat_dist} for the same choice of control gains and parameters. As can be verified, the error $s_2(t)$ still converges close to zero, leading $\hat{d}_1(t)$ to practically converge to a constant value and retaining the boundedness of the control input.



\section{Conclusion} \label{sec:conclusion}
We consider the asymptotic regulation problem for high-order MIMO nonlinear control-affine systems with unknown dynamic terms and control directions under state and input constraints.
We propose Barrier Integral Control to confine the state in a pre-defined funnel and guarantee its asymptotic convergence to zero from all initial conditions. An extension using reference modification accounts for input-saturation constraints. Future efforts will be devoted towards extending the proposed scheme to tracking of arbitrary trajectories and systems with
unmatched uncertainties.

{
\section*{Appendix: Proof of Lemma \ref{lemma:s bounds}}}

{
By integrating $\dot{s}_{k-1_j} =-\lambda s_{k-1_j} + s_{k_j}$ and using $|s_{k_j}| \leq \bar{s}$, we obtain 
$|s_{k-1_j}(t)| \leq |s_{k-1_j}(0)|\exp(-\lambda t) + \frac{\bar{s}}{\lambda}$. By proceeding in a recursive manner, we obtain 
\begin{align*}
|s_{i_j}(t)| \leq \exp(-\lambda t) \sum_{ \ell=i}^{k-1} |s_{\ell_j}(0)| \frac{t^{\ell-i}}{(\ell-i)!} + \frac{\bar{s}}{\lambda^{k-i}}
\end{align*}
for all $t\geq 0$, $i\in\{1,\dots,k-1\}$, $j\in\{1,\dots,n\}$. By further noting that $t^m \exp(-\lambda t) \leq \frac{m^m}{\epsilon^m \exp(1)^m} \exp(-(\lambda - \epsilon)t)$ for any positive integer $m$ and $\epsilon \in (0,\lambda)$, we obtain   
\begin{align*}
& |s_{i_j}(t)| \leq  \bar{s}_i \exp(-\lambda_0 t) +\frac{\bar{s}}{\lambda^{k-i}}\\
& \bar{s}_i \coloneqq \sum_{ \ell=i}^{k-1}  \left(\frac{(\ell-i)}{\epsilon \exp(1)}\right)^{(\ell-i)} \frac{|s_{\ell_j}(0)|}{(\ell-i)!}   
\end{align*}
for all $t\geq 0$, $i\in\{1,\dots,k-1\}$, $j\in\{1,\dots,n\}$, where $\lambda_0 \coloneqq \lambda - \epsilon > 0$. By using \eqref{eq:s_i}, it can be shown recursively that $|e_{1_j}(t)| = |s_{1_j}(t)| \leq \bar{e}_{1,1}\exp(-\lambda_0 t) + \bar{e}_{1,2} \frac{\bar{s}}{\lambda^{k-1}} $, with $\bar{e}_{1,1} = \bar{s}_1$,  $\bar{e}_{1,2} = 1$, and 
\begin{align*}
|e_{i_j}(t)| \leq \bar{e}_{i,1} \exp(-\lambda_0 t) + \bar{e}_{i,2}  \frac{\bar{s}}{\lambda^{k-i}},
\end{align*}
for all $t\geq 0$, $i\in\{2,\dots,k-1\}$, $j\in\{1,\dots,n\}$, 
with $\bar{e}_{i,1} \coloneqq \bar{s}_i + \sum_{\ell=1}^{i-1} \left( \begin{matrix}
i- 1\\
\ell
\end{matrix} \right)
\lambda^\ell \bar{e}_{i-\ell,1}$ 
and $\bar{e}_{i,2} \coloneqq 1 +  \sum_{\ell=1}^{i-1} \left( \begin{matrix}
i- 1\\
\ell
\end{matrix} \right)$.
}
\bibliographystyle{unsrt}        
\bibliography{references}           



\end{document}